\documentclass[10pt, conference, compsocconf]{IEEEtran}
\usepackage[T1]{fontenc}
\usepackage[latin9]{inputenc} 
\usepackage{verbatim}
\usepackage{textcomp}
\usepackage{amsthm}
\usepackage{amsmath}
\usepackage{graphicx}
\usepackage{amssymb} 
\usepackage[english]{babel}

\IEEEoverridecommandlockouts

\makeatletter

\newcommand{\lyxmathsym}[1]{\ifmmode\begingroup\def\b@ld{bold}
  \text{\ifx\math@version\b@ld\bfseries\fi#1}\endgroup\else#1\fi}

\author{\IEEEauthorblockN{Jacek Sroka\thanks{This work was sponsored by the Polish National Science Centre.}}
\IEEEauthorblockA{University of Warsaw, Poland\\
Email: {\tt sroka@mimuw.edu.pl}}
\and
\IEEEauthorblockN{Jan Hidders}
\IEEEauthorblockA{Delft University of Technology,
The Netherlands\\
Email: {\tt a.j.h.hidders@tudelft.nl}}
}

\theoremstyle{plain}
\newtheorem{thm}{Theorem}
  \theoremstyle{definition}
  \newtheorem{defn}[thm]{Definition}
  \theoremstyle{plain}
  
  \theoremstyle{plain}
  \newtheorem{lem}[thm]{Lemma}
  \newtheorem{prop}[thm]{Proposition}
  \theoremstyle{plain}
  \newtheorem{cor}[thm]{Corollary}

\makeatother

\usepackage{babel}

\newcommand{\tc}[1]{\mathbf{tc}(#1)}
\newcommand{\pc}[1]{\mathbf{pc}(#1)}

\begin{document}

\title{On Generating {*}-Sound Nets with Substitution}
\maketitle
\begin{abstract}
We present a method for hierarchically generating sound workflow nets by substitution of nets with multiple inputs and outputs. We show that this method is correct and generalizes the class of nets generated by other hierarchical approaches. The method involves a new notion of soundness which is preserved by the generalized type of substitution that is presented in this paper. We show that this notion is better suited than {*}-soundness for use with the presented type of generalized substitution, since {*}-soundness is not preserved by it. It is moreover shown that it is in some sense the optimal notion of soundness for the purpose of generating sound nets by the presented type of substitution.
\end{abstract}

\section{Introduction}

Among all the different formalisms for modelling processes, Petri nets \cite{Reisig:2008} offer the distinct benefits of combining an easy-to-understand visual notation with a large body of practical and theoretical work on efficient and effective reasoning over them. This has made them very popular for modelling of and reasoning over complex systems and specifically business processes and business workflows.

\begin{figure}
\begin{centering}
\includegraphics[clip,width=0.4\textwidth]{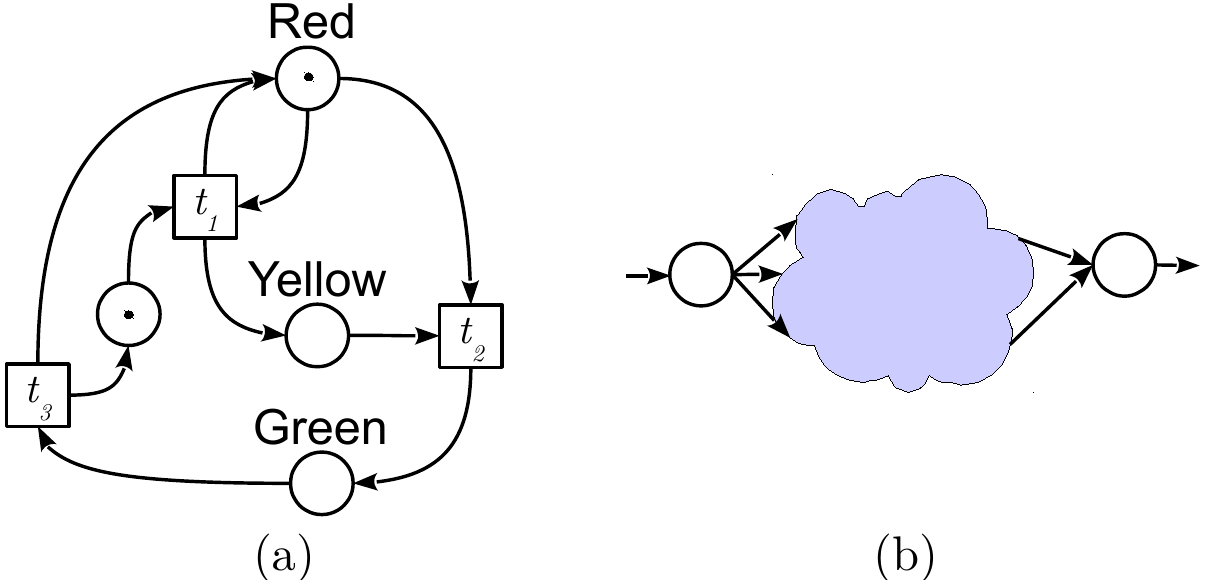}
\par\end{centering}
\caption{\label{fig:petri_net_examples}a) an example Petri net simulating the German traffic lights and b) workflow nets is a special kind of Petri net}
\end{figure}

An example of a Petri net modelling the German traffic lights is presented in Figure~\ref{fig:petri_net_examples} (a). It is composed of two kinds of nodes:  circular places and rectangular transitions. Places can store tokens, depicted by black dots, that represent availability of some resource or occurrence of some condition. The transitions are the active components that consume tokens from their input places and produce tokens into their output places. Input places of a transition are those that are connected by an edge leading from the place to the transition, while the output places are those that are connected by an edge leading from the transition to the place. In the net from the example there is one token in the top place representing a red light and another token in the leftmost place preventing the transition $t_1$ from firing multiple times in a row. In such a state only $t_1$ is active, i.e., there are tokens in each of its input places. When it fires, it consumes tokens from all its input places, i.e., the top and the leftmost place, and produces a token into each of its output places, i.e., the top place and the middle place. The resulting state represents red and yellow lights turned on simultaneously. Then, only the transition $t_2$ will be enabled. After it fires, the net will reach a state with only one token in the bottom place, which represents the green light being on and all the other places being empty. Then, only $t_3$ will be enabled and when it fires, the initial state from the figure is recreated.

For describing business processes and business workflows a specific class of Petri nets, called workflow nets, was introduced which features attractive modelling and analytic properties such as easy-to-verify notions of correctness. Workflow nets are Petri nets with one "global" input and one "global" output place (see Figure~\ref{fig:petri_net_examples} (b)), representing the beginning and the end of the flow, respectively, that become strongly connected when we add a transition from the output place to the input place. The workflow starts with one token in the input place and while the workflow is running, it follows the usual firing rules of a Petri net. It is assumed to have completed correctly, when the net reaches a state with exactly one token, which is placed in the output place.

The correctness of any model, and therefore also workflow nets, ultimately depends on whether it correctly models the domain in question. However, certain properties such as the absence of livelocks, deadlocks, and other anomalies are desirable and can be checked independently of the specific domain. Among these the \emph{soundness} property of the workflow net is considered the most important. This notion was originally proposed by van der Aalst in~\cite{Aalst1998workflow} and since then several alternative notions of soundness have been proposed and studied. Informally speaking soundness means two things. First, that if we start with an initial token in the input place, then no matter how we proceed with the execution of the workflow, we can always end up in the final state with one token in the output place. Second, that every subtask can be potentially executed, i.e., there is at least one correct run of the workflow net in which this subtask is executed. An overview of the research on soundness of workflow nets with additional decidability results can be found in~\cite{journals/fac/AalstHHSVVW11}.

\begin{figure}
\begin{centering}
\includegraphics[width=0.42\textwidth]{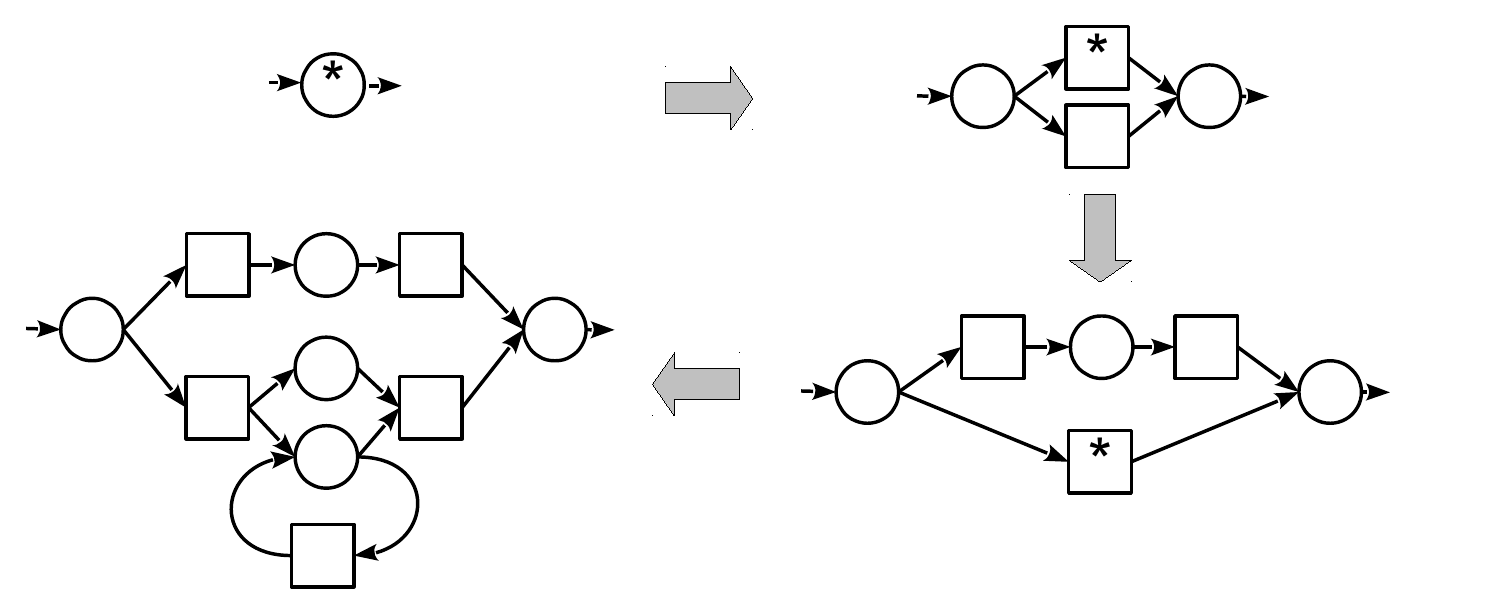}
\par\end{centering}
\caption{\label{fig:refinement-example}An example of a top-down construction of a workflow net}
\end{figure}

Process modellers have a choice between two main approaches if they want to produce sound workflow nets. The first is to design the workflow as they like and then use the different existing algorithms to determine if the desired requirements are met, like in \cite{kSound,reviewer2a,reviewer2b} or \cite{conf/apn/EsparzaS89a}. The second is to construct the design step by step and use only manipulations and combinations of nets that are guaranteed to produce sound nets \cite{Desel:2005:FCP:1205779}. In this paper we investigate the second approach and in particular focus on a structural approach where the net is constructed in a top-down fashion. This means that the system is designed by first specifying a workflow net that provides a high-level description of the process by summarizing it at a high abstraction level in terms of high-level actions, and then refining this workflow net in a stepwise fashion by replacing nodes that represent high-level actions with workflow nets which describe these actions in more detail. An example of such a top-down construction is given in Figure~\ref{fig:refinement-example}. At each step a certain node, marked by a *, is substituted with another workflow net. As is illustrated here the substituted net may also start with a transition, rather then a place. By restricting the type of nets we can start from and the type of nets we can substitute with, it can be guaranteed that the resulting net is always sound.

An additional advantage of such a hierarchical approach is that it produces workflow nets with an explicit and natural hierarchical structure, which considerably aids the understandability of the specification. It can be used in the design and analysis tools and allow the user to zoom in and out of specific parts of the net by either expanding or collapsing nodes according to the hierarchical structure. In addition the structure can often be matched with the organizational hierarchy of the organization that hosts the specified workflow, and therefore be linked with for example the levels of management. Moreover, the hierarchical structure can help with specifying elegantly the handling of exceptions and recovery from unexpected situations~\cite{wachtel2003,wachtel2006}. For a more elaborate motivation and description of the advantages of a hierarchical net design method the reader is referred to~\cite{DBLP:conf/caise/DumasRMMRS12, Polyvyanyy2012518}.

The specific refinement approach that we take in this paper works as follows. We always start with a simple type of net that we already know to have the desired soundness properties. Then we allow the substitution of a single node, either a place or a transition, with a workflow net that we also already know to possess the desired properties. We will show that for suitable soundness properties and specific types of substitutions it will hold that the soundness properties are preserved, i.e., the result of the substitution also has the soundness properties. This allows us to start from a small set of simple nets that are known to have the desired properties, and then generate from them a larger class of nets that also have these properties by closing the class under substitution, i.e., if two nets are in the class, then the substitution of one net into the other is also in the class. This idea of net refinements is quite old, and the first papers were published in the early 90's, like \cite{Brauer:1991kn}. Methods for stepwise refinements were studied in numerous papers, including \cite{Suzuki198351}, \cite{10.1109/CSIE.2009.831}, \cite{Padberg200197} or \cite{huang_04_structure}. An approach that we will in particular focus on is the one presented by van Hee et al.\  in \cite{DBLP:conf/apn/HeeSV03} where two large classes of simple workflow nets, based on state machines and marked graphs, are identified which are readily observed to be sound, and then it is shown that when closed under substitution we get a larger class of workflow nets, called ST-nets, which contains also only sound workflow nets.

\begin{figure}
\begin{centering}
\includegraphics[width=0.32\textwidth]{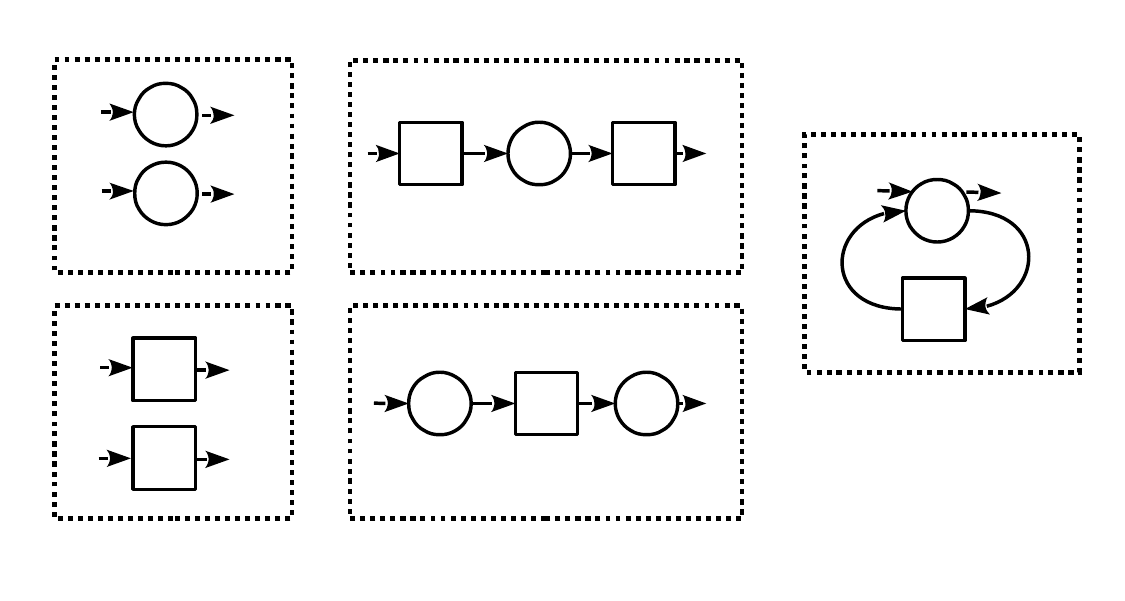}
\par\end{centering}
\caption{\label{fig:hierarch-subst-nets} Allowed substitution nets for generating hierarchical nets}
\end{figure}

Another approach for generating sound nets by substitution that can be found in the literature works as follows. We always start with a net consisting of a single place, and allow only the substitution with one of a finite set of simple nets as for example those shown in Figure~\ref{fig:hierarch-subst-nets}. Strictly speaking these are not workflow nets since they are are allowed to have multiple input and output nodes. When such a net is substituted, each input and output node is connected to the surrounding net in the same way as it would have been if it was the only input or output node. This approach is taken by Wachtel et al.\ in in \cite{wachtel2003} and van Hee et al.\  in \cite{hee2009} and the class of nets that can be generated this way is referred to as the class of \emph{hierarchical nets}. Interestingly enough, this class is slightly different from the one generated by the approach in \cite{DBLP:conf/apn/HeeSV03}, and is neither strictly larger nor smaller. It is the main goal of this paper to investigate the combination of these two approaches and see if it allows the generation of even larger classes of sound nets.

\begin{figure}
\begin{centering}
\includegraphics[clip,width=0.42\textwidth]{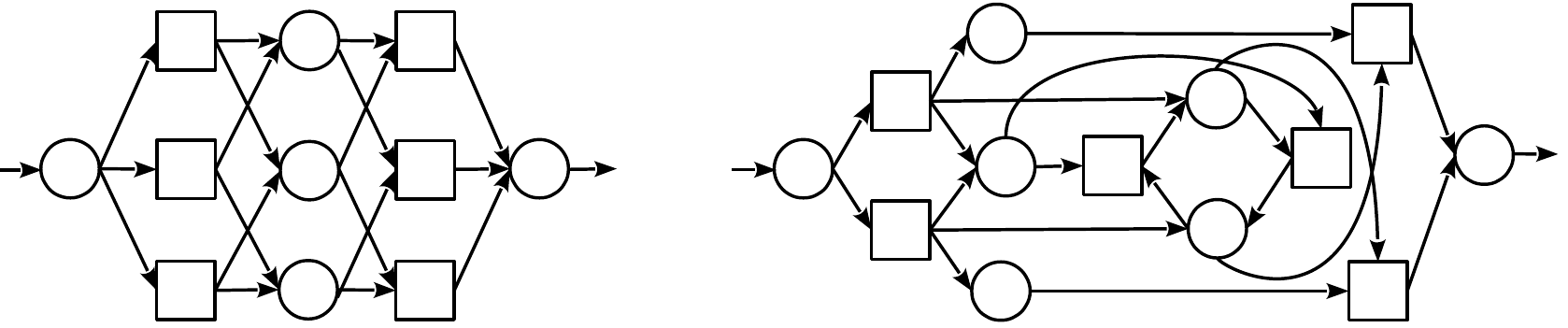}
\par\end{centering}
\caption{\label{fig:1-sound-examples} Examples of nets that are sound in the classical sense but have problems when initiated with 2 tokens, i.e., are $1$-sound but not $2$-sound}
\end{figure}

For the approach chosen in this paper we need a special notion of soundness. This is because, as was observed by van Hee et al. in \cite{DBLP:conf/apn/HeeSV03}, it is unfortunately in general not true that soundness as defined earlier is preserved by substitution, i.e., if we substitute a sound net in another sound net the result is not necessarily sound. This is related to the fact that although if we execute a sound workflow, starting with a single token, then we will end up with a single token in the output place and no other tokens anywhere, it could be that if we start the same workflow with 2 tokens, it does not necessarily mean that the final marking will have 2 tokens in the output place. It can therefore happen that substitution of such a workflow net will lead to an unsound net. Two classical examples of such nets, which are sound in classical sense, but have problems when initiated with 2 tokens, are presented after~\cite{DBLP:conf/apn/HeeSV03,journals/fac/AalstHHSVVW11} in Figure~\ref{fig:1-sound-examples}.

For this reason the notion of $k$-soundness was introduced by van Hee et al, where $k$ is a parameter for which whenever we start with $k$ tokens, the net will end without deadlock having exactly $k$ tokens in the output place, while all other places will be unmarked. It was proven that $k$-soundness forms a strict hierarchy, which means that for every $k$ there exist a workflow net which is $k$-sound and not $(k+1)$-sound. The nets in Figure~\ref{fig:1-sound-examples} are $1$-sound but not $2$-sound. The notion of {*}-soundness holds for nets, which are sound for every $k$. It is shown by van Hee et al.\ in \cite{DBLP:conf/apn/HeeSV03}, that this type of soundness is preserved by substitution for their kind of nets. In the same paper van Hee et al. define a large class of nets by starting from very simple classes that are syntactically easy to identify and can be straightforwardly shown to be {*}-sound, and then generating more {*}-sound nets by substitution.

Since in this paper we consider a more generalized notion of substitution that also allows substitution of nets with multiple input and output places and allows flow edges that arrive in input places and leave from output places, our approach requires a slightly generalized notion of soundness that we call \emph{substitution soundness} and which is indeed preserved by the generalized type of substitution that we propose.

The structure of the paper is as follows. After introducing the notions of a Petri net, workflow net and soundness we propose a new classes of nets, called p-WF nets and t-WF nets. Informally such nets have the border nodes being places or transitions respectively. \emph{AND-OR} nets being special classes of p-WF nets and t-WF nets are introduced in Section~\ref{sec:AND-OR_nets}. We make some remarks on their properties and specify how the substitutions are performed. Next, we address the problem of soundness preservation during substitution in Section~\ref{sec:subsoundness} and introduce the notion of \emph{substitution-soundness} (\emph{sub-soundness} for short). The main two theorems of this section state that soundness is preserved when a sub-sound t-WF net is substituted for a transition of a sub-sound p-WF net or t-WF net and when a sub-sound p-WF net is substituted for a place of a sub-sound p-WF net or t-WF net. In Section~\ref{sec:subsoundness_AND-OR} we prove that the introduced AND-OR nets are sub-sound in general. 

A preliminary version of this paper was presented in 2011 at the $11^{th}$ International Conference on Application of Concurrency to System Design in Newcastle upon Tyne, United Kingdom, see~\cite{10.1109/ACSD.2011.26}. Apart of providing a more elaborate discussion of the results, the main extensions in this paper include the complete versions of proofs and discussion on whether substitution soundness is the right notion of soundness, i.e., is the weakest condition necessary for constructing nets by refinement. 

\section{Basic Terminology}\label{sec:terminology}

Let $S$ be a set. A bag (multiset) $m$ over $S$ is a function $m:S\rightarrow\mathbb{N}$. We use $+$ and $\lyxmathsym{\textminus}$ for the sum and the difference of two bags and $=$, $<$, $>$, $\le$, $\ge$ for comparisons of bags, which are defined in a standard way. We overload the set notation, writing $\emptyset$ for the empty bag and $\in$ for the element inclusion. We list elements of bags between brackets, e.g. $m=[p^{2},q]$ for a bag $m$ with $m(p)=2$, $m(q)=1$, and $m(x)=0$ for all $x\notin\{p,q\}$. The shorthand notation $k.m$ is used to denote the sum of $k$ bags $m$. The size of a bag $m$ over $S$ is defined as $|m|=\Sigma_{s\in S}m(s)$.
\begin{defn}
[Petri net] A \emph{Petri net} is a tuple $N=(P,T,F)$ with $P$ a finite set of places, $T$ a finite set of transitions such that $P\cap T=\emptyset$ and $F\subseteq(T\times P)\cup(P \times T)$ the set of flow edges.
\end{defn}
A path of a net is a non-empty sequence $(x_{1},...,x_{n})$ of nodes where for all $i$ such that $1\leq i\leq n-1$ it holds that $(x_{i},x_{i+1})\in F$. Markings are states (configurations) of a net and the set of markings of $N=(P,T,F)$ is the set of all bags over $P$ and denoted as $\mathbf{M_N}$. Given a transition $t\in T$, the preset $\bullet t$ and the postset $t \bullet$ of $t$ are the sets $\{p\mid (p,t) \in F\}$ and $\{p\mid (t,p) \in F\}$, respectively. Analogously we write $\bullet p$, $p \bullet$ for pre- and postsets of places. To emphasize the fact that the preset/postset is considered within some net $N$, we write $\bullet_N a$ , $a \bullet_N$. We overload this notation further allowing to apply preset and postset operations to a set $B$ of places/transitions, which is defined as the union of pre-/postsets of elements of $B$. A transition $t \in T$ is said to be enabled in marking $m$ iff $\bullet t \le m$. For a net $N=(P,T,F)$ with markings $m_1$ and $m_2$ and a transition $t \in T$ we write $m_1 \stackrel{t}{\longrightarrow}_N m_2$, if $t$ is enabled in $m_1$ and $m_2 = m_1 - \bullet t + t \bullet$. For a sequence of transitions $\sigma=( t_1,\ldots, t_n )$ we write $m_1\stackrel{\sigma}{\longrightarrow}_N m_n$, if $m_1 \stackrel{t_1}{\longrightarrow}_N m_2 \stackrel{t_2}{\longrightarrow}_N \ldots \stackrel{t_n}{\longrightarrow}_N m_n$, and we write $m_1 \stackrel{*}{\longrightarrow}_N m_n$, if there exists such a sequence $\sigma\in T^{*}$. We will write $m_1 \stackrel{t}{\longrightarrow} m_2$, $m_1 \stackrel{\sigma}{\longrightarrow} m_n$ and $m_{1}\stackrel{*}{\longrightarrow}m_{n}$, if $N$ is clear from the context. 

\begin{figure}
\begin{centering}
\includegraphics[clip,width=0.37\textwidth]{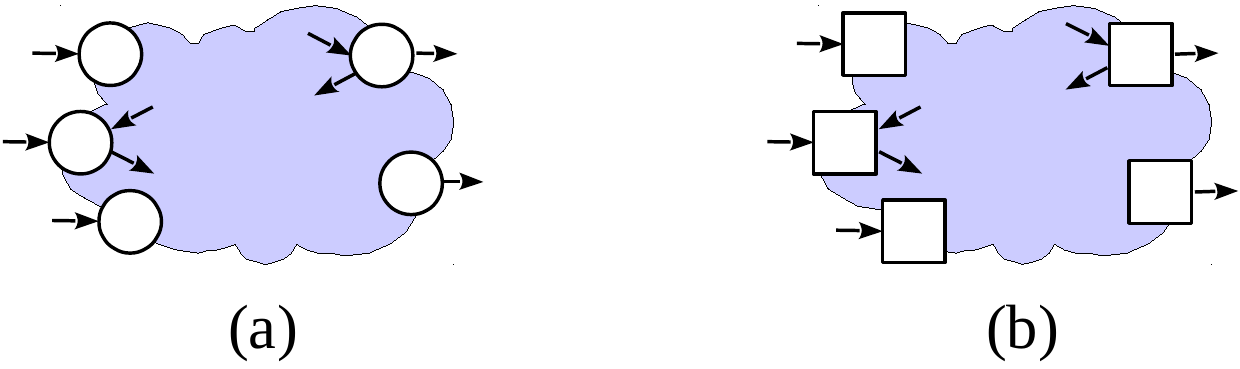}
\par\end{centering}
\caption{\label{fig:wf_net_examples}A generalized notion of workflow net: a) pWF net and b) tWF net}
\end{figure}

We generalize the usual notion of workflow net as introduced by van der Aalst in \cite{Aalst1998workflow} by allowing multiple input and output places, allowing transitions as input and output nodes and also allowing input nodes to have incoming edges and output nodes to have outgoing edges (see Figure~\ref{fig:wf_net_examples}).

\begin{figure}
\begin{centering}
\includegraphics[clip,width=0.45\textwidth]{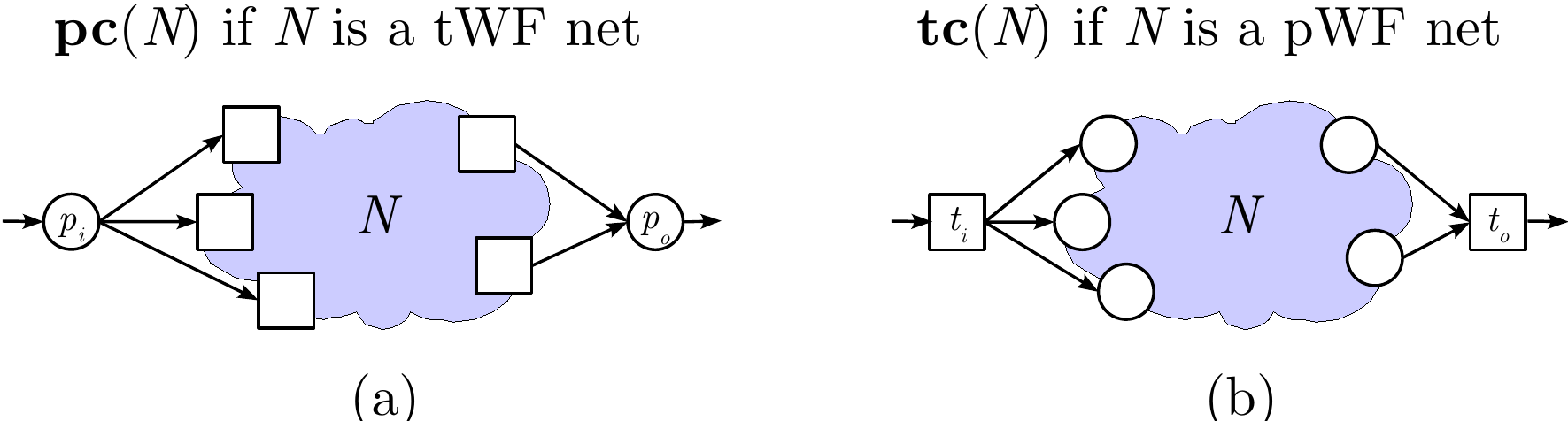}
\par\end{centering}
\caption{\label{fig:place-compl}The place completion of a tWF net and a transition completion of a pWF net}
\end{figure}

\begin{defn}
[Workflow net] A \emph{place Workflow net} (pWF net) is a tuple $N=(P,T,F,I,O)$ where $(P,T,F)$ is a Petri net with a non-empty set $I\subseteq P$ of \emph{input places} and a non-empty set $O\subseteq P$ of \emph{output places} such that (1) every node in $P\cup T$ is reachable by a path from at least one node in $I$ and (2) from every node in $P\cup T$ we can reach at least one node in $O$. A \emph{transition Workflow net} (tWF net) is similar to a place Workflow net except that $I$ and $O$ are non-empty subsets of $T$. A \emph{workflow net} (WF net) is either a pWF net or tWF net. 
\end{defn}
\emph{A} workflow net is called a \emph{one-input} workflow net if $I$ contains one element, and a \emph{one-output} workflow net if $O$ contains one element. In \cite{Aalst1998workflow} workflow nets are restricted to one-input one-output place Workflow nets. We generalize this but define for all workflow nets the corresponding one-input one-output pWF net as follows. The \emph{place-completion} of a tWF net $N=(P,T,F,I,O)$ is denoted as $\pc{N}$ and is a one-input one-output pWF net that is constructed from $N$ by adding places $p_{i}$ and $p_{o}$ such that $p_{i}\bullet=I$ and $\bullet p_{o}=O$ and setting the input set and output set as $\{p_{i}\}$ and $\{p_{o}\}$, respectively. This is illustrated in Figure~\ref{fig:place-compl} (a). Note that we distinguish $I$ nodes with half unconnected incoming arrows and $O$ nodes with half unconnected outgoing arrow. The \emph{transition-completion} of a pWF net $N=(P,T,F,I,O)$ is denoted as $\tc{N}$ and is a one-input one-output tWF net that is constructed from $N$ by adding transitions $t_{i}$ and $t_{o}$ such that $t_{i}\bullet=I$ and $\bullet t_{o}=O$ and setting the input set and output set as $\{t_{i}\}$ and $\{t_{o}\}$, respectively. This is illustrated in Figure~\ref{fig:place-compl} (b). 

We will focus in this paper on a particular kind of soundness, namely the soundness that guarantees the reachability of a proper final state. We generalize this for the case where there can be more than one input place and these contain one or more tokens in the initial marking. We also provide a generalization of soundness for tWF nets, which intuitively states that, if in total there are $k$ firings of input transitions, then the computation will end in an empty marking after in total $k$ firings of the output transitions.
\begin{defn}
[$k$ and *-soundness] A pWF net $N=(P,T,F,I,O)$ is said to be \emph{$k$-sound} if for each marking $m$ such that $k.I\stackrel{*}{\longrightarrow}m$ it holds that $m\stackrel{*}{\longrightarrow}k.O$. We call $N$ \emph{{*}-sound} if it is $k$-sound for all $k\geq1$. We say that these properties hold for tWF net $N$ if they hold for $\pc{N}$. 
\end{defn}

It would be nice if transition-completion would not affect the {*}-soundness of a net just like place-completion does (by definition). However this is only partially true as is shown in the following theorem.

\begin{thm}
\label{thm:tc-soundness}Every pWF net $N$ is {*}-sound if $\tc{N}$ is {*}-sound but not vice versa.
\end{thm}

\begin{figure}
\begin{centering}
\includegraphics[clip,width=0.3\textwidth]{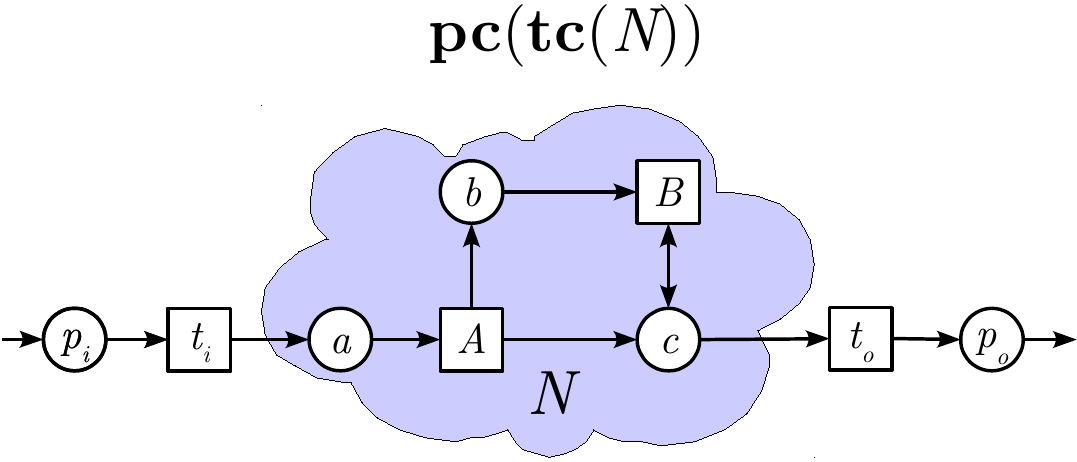}
\par\end{centering}
\caption{\label{fig:pc-tc-soundness}A counterexample showing that {*}-soundness is not preserved by transition completion and also not under substitution}
\end{figure}

\begin{IEEEproof}
Let $N=(P,T,F,I,O)$ and $N'=\pc{\tc{N}}=(P',T',F',I',O')$ with $t_i$ and $t_o$ being the added input and output transitions of $\tc{N}$, respectively. Recall that by definition $\tc{N}$ is {*}-sound iff $\pc{\tc{N}}$ is {*}-sound. We assume that $\tc{N}$ is {*}-sound, that is $N'$ is {*}-sound. Observe that $k.I'\stackrel{*}{\longrightarrow}_{N'}k.I$ by letting input transitions $t_{i}$ of $\tc{N}$ fire $k$ times. Assume that $k.I\stackrel{*}{\longrightarrow}_{N}m$. Since $N$ is embedded in $N'$, it then follows that $k.I'\stackrel{*}{\longrightarrow}_{N'}m$. From the {*}-soundness of $N'$ it follows that $m\stackrel{\sigma'}{\longrightarrow}_{N'}k.O'$ for some $\sigma'\in (T')^{*}$. However, we can omit the firings of $t_{o}$ from $\sigma'$ and obtain $\sigma$ such that $m\stackrel{\sigma}{\longrightarrow}_{N'}k.O$. Since $\sigma$ cannot contain $t_{i}$ it follows that $m\stackrel{\sigma}{\longrightarrow}_{N}k.O$ and therefore $m\stackrel{*}{\longrightarrow}_{N}k.O$.

The counterexample in Figure~\ref{fig:pc-tc-soundness} shows that not for every {*}-sound pWF net $N$ it holds that $\tc{N}$ is {*}-sound. Observe that $N$ is {*}-sound. However, the shown $\pc{\tc{N}}$ is not since from the marking $[p_{i}]$ it can reach $[b,c]$ and therefore $[b,p_{o}]$ after which no transition is enabled. Since  $\pc{\tc{N}}$ is not 1-sound, then by definition $\tc{N}$ is also not 1-sound and thus not {*}-sound.
\end{IEEEproof}

\section{AND-OR nets}\label{sec:AND-OR_nets}

\begin{figure}
\begin{centering}
\includegraphics[width=0.35\textwidth]{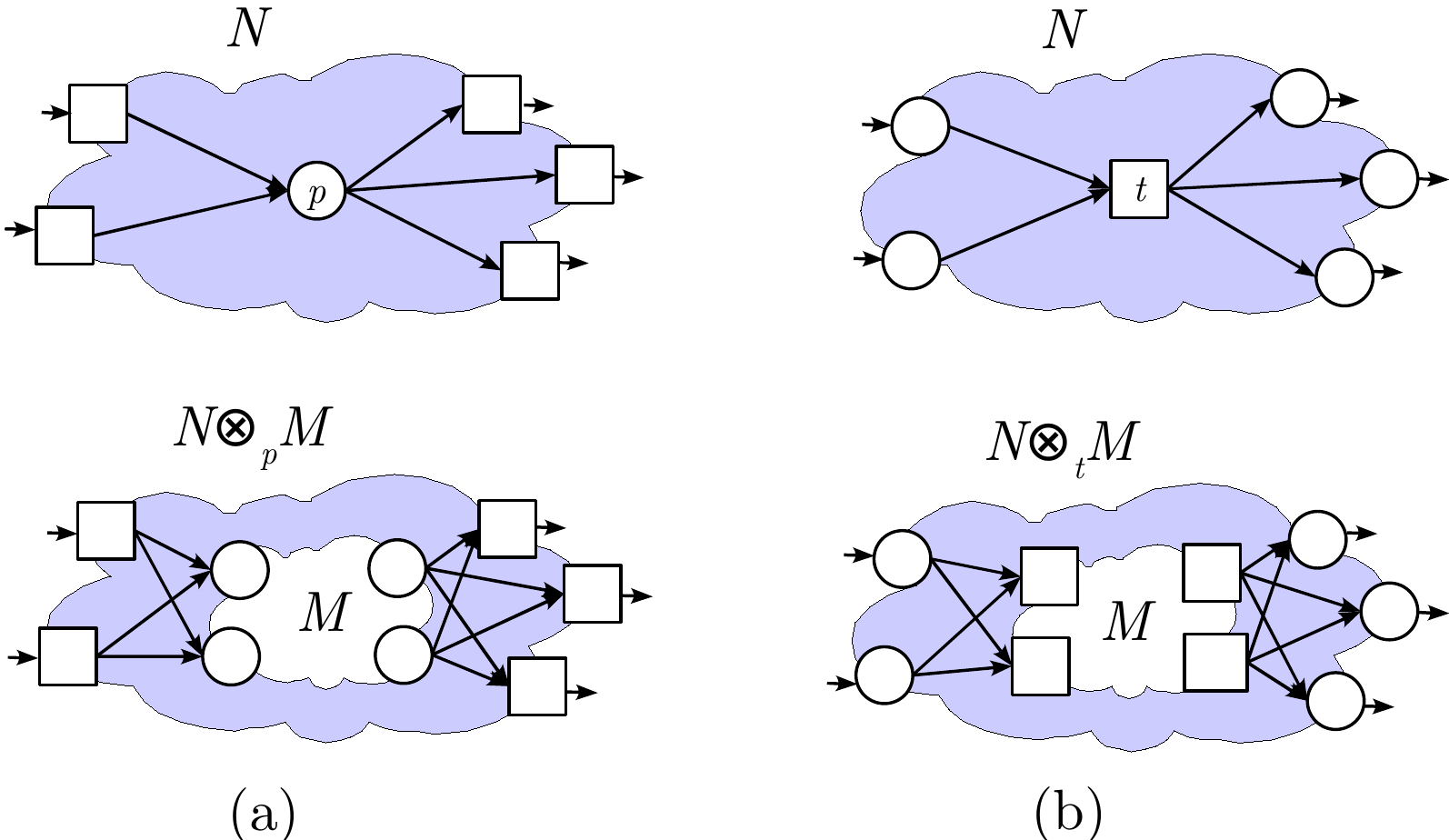}
\par\end{centering}
\caption{\label{fig:subst-illustr}Illustration of place substitution and transition substitution}
\end{figure}

To generate a large class of nets we will consider general substitutions where places and transitions are replaced with pWF nets and tWF nets, respectively. We introduce for this purpose a notion of substitution that is based on the one introduced by van Hee et al. in~\cite{DBLP:conf/apn/HeeSV03} but generalized so it can substitute nets with multiple input nodes and multiple output nodes.
\begin{defn}
[Place substitution, Transition substitution] Consider two \emph{disjoint} WF nets $N$ and $M$, i.e., if $N=(P,T,F,I,O)$ and $M=(P',T',F',I',O')$, then $(P\cup T)\cap(P'\cup T')=\emptyset$. 

\emph{Place substitution:} If $p$ is a place in $N$ and $M$ is a pWF net, then we define the result of substituting $p$ in $N$ with $M$, denoted as $N\otimes_{p}M$, as the net that is obtained if in $N$ we remove $p$ and the edges in which it participates and replace it with the net $M$ and edges such that $\bullet p'=\bullet p$ for each input place $p' \in I'$ of $M$ and $p'\bullet=p\bullet$ for each output place $p' \in O'$ of $M$. If $p\in I$ then $p$ is replaced in the set of input nodes of the resulting net with $I'$, i.e., the input set of $N\otimes_{p}M$ is $\left( I \setminus \{ p \} \right) \cup I'$, and if $p\in O$ then $p$ is replaced in the set of output nodes of the resulting net with $O'$, i.e., the output set of $N\otimes_{p}M$ is $\left( O \setminus \{ p \} \right) \cup O'$. Otherwise, the input and output sets of $N\otimes_{p}M$ are the same as the respective sets for $N$.

\emph{Transition substitution:} Likewise, if $t$ is a transition in $N$ and $M$ is a tWF net, then we define the result of substituting $t$ in $N$ with $M$, denoted as $N\otimes_{t}M$, as the net that is obtained if in $N$ we remove $t$ and the edges in which is participates and replace it with the net $M$ and edges such that $\bullet t'=\bullet t$ for each input transition $t' \in I'$ of $M$ and $t'\bullet=t\bullet$ for each output transition $t' \in O'$ of $M$. If $t\in I$ then $t$ is replaced in the set of input nodes of the resulting net with $I'$, i.e., the input set of $N\otimes_{t}M$ is $\left( I \setminus \{ t \} \right) \cup I'$, and if $t\in O$ then $t$ is replaced in the set of output nodes of the resulting net with $O'$, i.e., the output set of $N\otimes_{t}M$ is $\left( O \setminus \{ t \} \right) \cup O'$. Otherwise, the input and output sets of $N\otimes_{t}M$ are the same as the respective sets for $N$.
\end{defn}
The results of a place substitution and transition substitution are illustrated in Figure~\ref{fig:subst-illustr} (a) and (b), respectively. It is not hard to see that if $N$ and $M$ are WF nets and $n$ a node in $N$ then $N\otimes_{n}M$ is again a WF net. It also holds for all WF nets $A$, $B$ and $C$ that $(A\otimes_{a}B)\otimes_{b}C=A\otimes_{a}(B\otimes_{b}C)$ if $b$ is a node in $B$, and $(A\otimes_{a}B)\otimes_{b}C=(A\otimes_{a}C)\otimes_{b}B$ if $a$ and $b$ are nodes in $A$.

We will generate nets by starting from some basic classes of nets and allowing substitutions of places with pWF nets and transitions with tWF nets.
\begin{defn}
[Substitution closure] Given a class $C$ of nets we defined the substitution closure of $C$, denoted as $\mathbf{S}(C)$, as the smallest superclass of $C$ that is closed under transition substitution and place substitution, i.e., the following two rules hold: if $N$ and $M$ are disjoint nets in $\mathbf{S}(C)$ then (1) if $M$ is a pWF net and $p$ a place in $N$ then $N\otimes_{p}M$ is a net in $\mathbf{S}(C)$ and (2) if $M$ is a tWF net and $t$ a transition in $N$ then $N\otimes_{t}M$ is a net in $\mathbf{S}(C)$.
\end{defn}

As the basic nets with which we will start the generation process we will consider the nets that we call pAND nets, tAND nets, pOR nets and tOR nets, which are all illustrated in Figure~\ref{fig:And-or-net} with input and output nodes on the left-hand side and right-hand side, respectively. Informally we can describe AND nets as acyclic nets that consist only of AND splits and AND joins, and OR nets can be described as possibly cyclic nets consisting of only OR splits and OR joins. AND and OR nets are generalizations of marked graph/T-nets and state machines/S-nets~\cite{Desel:2005:FCP:1205779}, respectively, which both are restricted to exactly one input and output node. More formally, the AND and OR nets are defined as follows.

\begin{center}
\begin{figure}
\begin{centering}
\includegraphics[width=0.45\textwidth]{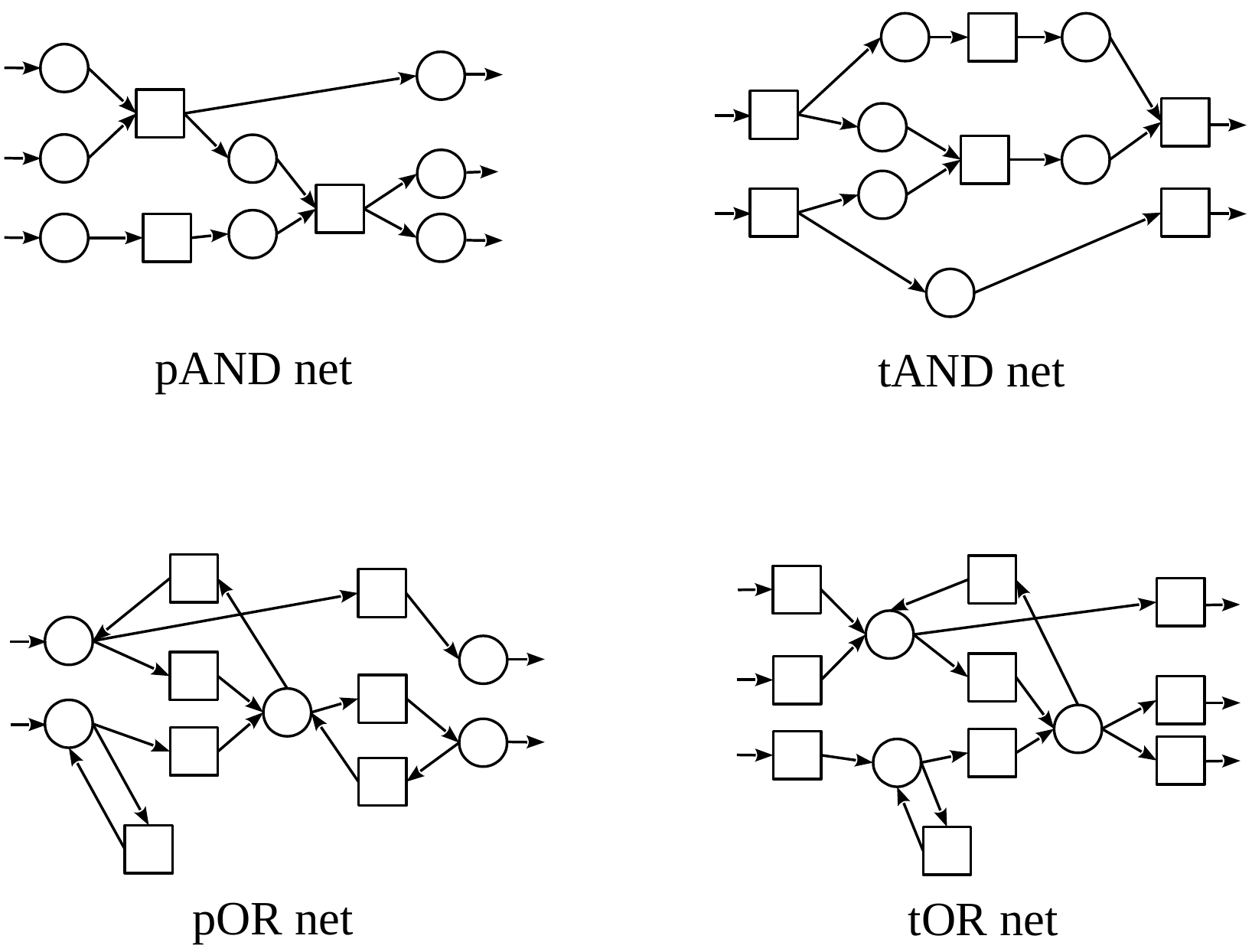}
\par\end{centering}
\caption{\label{fig:And-or-net}Examples of a pAND, tAND, pOR and tOR nets}
\end{figure}
\par\end{center}

\begin{defn}
[AND net] An \emph{AND net} is an acyclic WF net $(P,T,F,I,O)$ such that for every place $p\in P$ it holds that (1) $p\in I\wedge|\bullet p|=0$ or $p\notin I\wedge|\bullet p|=1$ and (2) $p\in O\wedge|p\bullet|=0$ or $p\notin O\wedge|p\bullet|=1$. An AND net that is a pWF net is called a pAND net, and if it is a tWF net it is called a tAND net. 
\end{defn}
OR nets are the counterpart of AND nets and are defined as follows.
\begin{defn}
[OR net] An \emph{OR net} is a WF net $(P,T,F,I,O)$ such that for every transition $t\in T$ it holds that (1) $t\in I\wedge|\bullet t|=0$ or $t\notin I\wedge|\bullet t|=1$ and (2) $t\in O\wedge|t\bullet|=0$ or $t\notin O\wedge|t\bullet|=1$. An OR net that is a pWF net is called a pOR net, and if it is a tWF net it is called a tOR net.
\end{defn}
\begin{center}

\begin{figure}
\begin{centering}
\includegraphics[clip,width=0.42\textwidth]{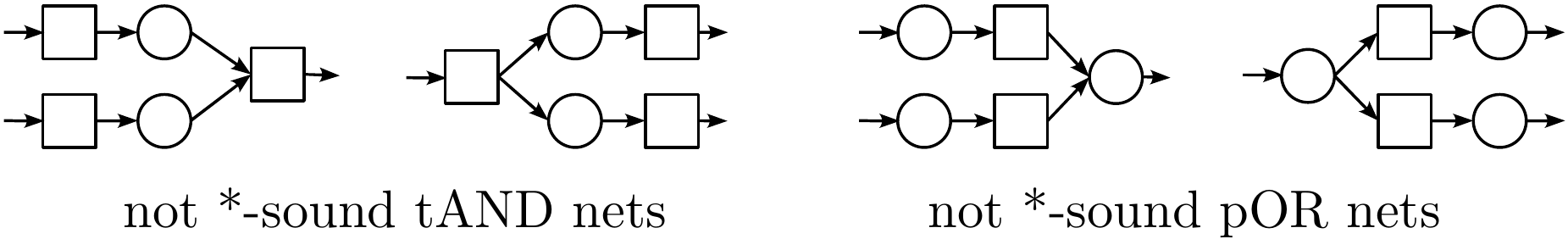}
\par\end{centering}
\caption{\label{fig:unsound-And-or-nets}Examples of tAND and pOR nets that are not {*}-sound}
\end{figure}
\par\end{center}

\begin{figure}
\begin{centering}
\includegraphics[clip,width=0.44\textwidth]{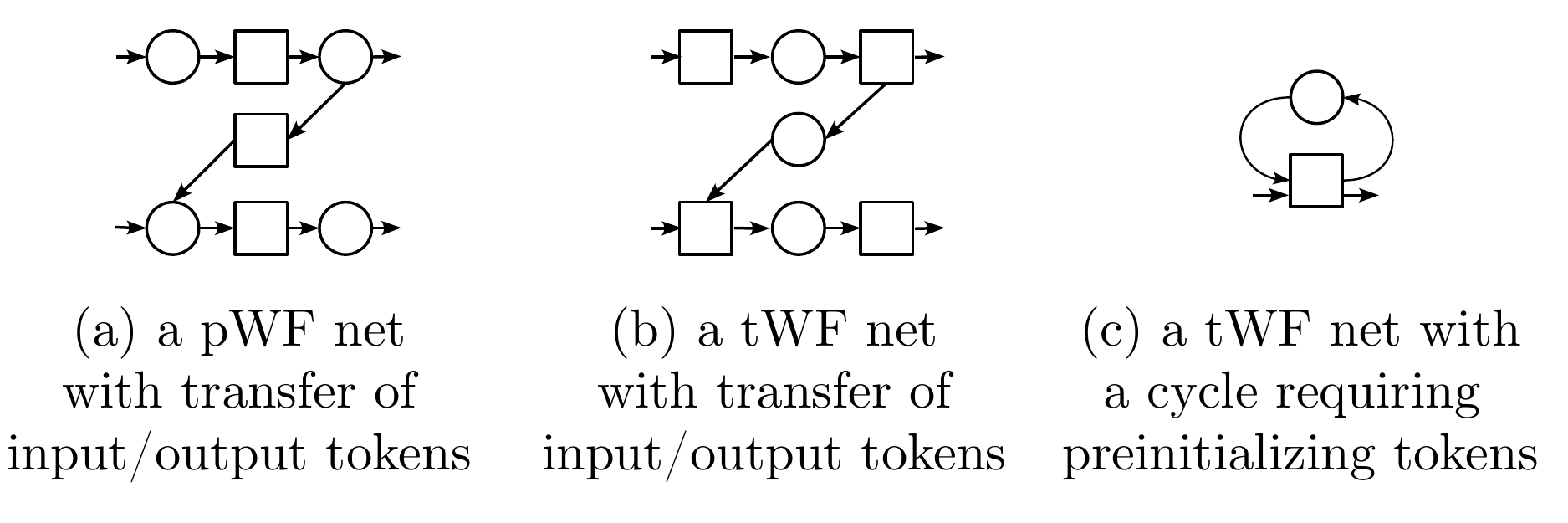}
\par\end{centering}
\caption{\label{fig:And-or-nets-forbidden-by-def}Unsound nets forbidden by the definition of AND and OR nets}
\end{figure}

Note that OR nets can contain cycles where AND nets by definition cannot, but otherwise they are each others dual. Also note that for the requirements over the edges, being an input node counts as having an input edge, and being an output node counts as having an output edge. To illustrate why this is so consider the nets (a) and (b) in Figure~\ref{fig:And-or-nets-forbidden-by-def}. In (a) we a see a pWF net that would be a pAND net if we ignored the requirement for input and output places. However, it will also be clear that this is not a 1-sound net since the token in the upper-right output place might be transferred to the lower-left input place, after which we cannot reach the final state. In (b) we see a tWF net that would be tOR net if we ignored the requirements for input and output places. Also here it is easy to see by looking at its place completion that this is not a 1-sound net. For tAND and pOR there are no such restrictions on the input and output nodes, since in AND nets the places are restricted and in OR nets the transitions are restricted. The requirement for acyclicity for AND nets is illustrated by the tWF net (c) in Figure~\ref{fig:And-or-nets-forbidden-by-def}. Clearly this net is not 1-sound since a run in which the transition fires requires an initial token in the place. However, its dual where the place is a transition and vice versa, is indeed 1-sound, which explains the asymmetry between AND and OR nets. 

For the AND and OR nets as defined here there are some straightforward soundness results in that all pAND and tOR nets are {*}-sound, and that for tAND and pOR nets this is the case if they are one-input one-output nets. The {*}-soundness of tOR nets follows from the {*}-soundness of ST-nets of van Hee et al. given by Theorem~17 in~\cite{DBLP:conf/apn/HeeSV03} and the definition of {*}-soundness for tWF nets by place completion. The {*}-soundness of pAND nets follows from Theorem~\ref{thm:tc-soundness} and the fact that for every pAND net $N$, its transition completion $\tc{N}$ is {*}-sound because $\pc{\tc{N}}$ is also an ST-net. Note that for this reasoning it is crucial that the place completion of a tOR net results in a net that is still an OR net, and that transition completion of a pAND net results in a net that is still an AND net. This is the case since in tOR nets and in pAND nets input nodes cannot have incoming edges and output nodes cannot have outgoing edges.

Observe that even without disallowing incoming edges for input nodes and outgoing edges for output nodes, a place completion of any tAND net results in an AND net and a transition completion of any pOR net results in OR net, i.e., results in nets that do not have AND splits/joins and OR splits/joins intermixed in a problematic way. Note also that for multi-input multi-output pOR nets an unsound transfer would be possible similarly as for pAND nets, but we limit the number of input/output places anyway. Finally, even though for tAND nets we do not limit the number of these edges in the definition, it follows from its acyclicity and existence of only one input and one output transition.

To understand the restriction to one-input one-output nets consider the examples of tAND and pOR nets in Figure~\ref{fig:unsound-And-or-nets} which are all nets with either multiple input nodes or multiple output nodes and which are all not {*}-sound. For the presented tAND net examples applying the place completion, which is required by the definition of soundness, would result in a net with AND splits/joins and OR splits/joins mixed in a wrong way. For the presented pOR net examples the problem originates from the nature of allowed OR splits/joins and the possibility of unequal numbers of input and output places. This is why, while generating nets with place and transition substitution, we limit ourselves to the following classes of nets: the class of pAND nets represented by \textbf{$\mathbf{pAND}$}, the class of one-input one-output tAND nets represented by $\mathbf{11tAND}$, the class of one-input one-output pOR nets represented by \textbf{$\mathbf{11pOR}$}, and the class of tOR nets represented by $\mathbf{tOR}$ (see Figure~\ref{fig:And-or-net11} for examples). For one-input one-output tAND nets the {*}-soundness follows immediately from the {*}-soundness of pAND nets because performing place completion of one-input one-output tWF nets does not create OR splits nor OR joins. For one-input one-output pOR nets we cannot refer to {*}-soundness of ST-nets, because they cannot have incoming edges for input places and outgoing edges for output places. Yet, by its construction the number of tokens in the net has to be constant and by reachability of input and output nodes in the definition of workflow net all tokens can be forced to reach the output place. Section~\ref{sec:subsoundness_AND-OR} provides formal proofs of stronger sub-soundness properties for all the basic classes discussed here.

\begin{center}
\begin{figure}
\begin{centering}
\includegraphics[width=0.35\textwidth]{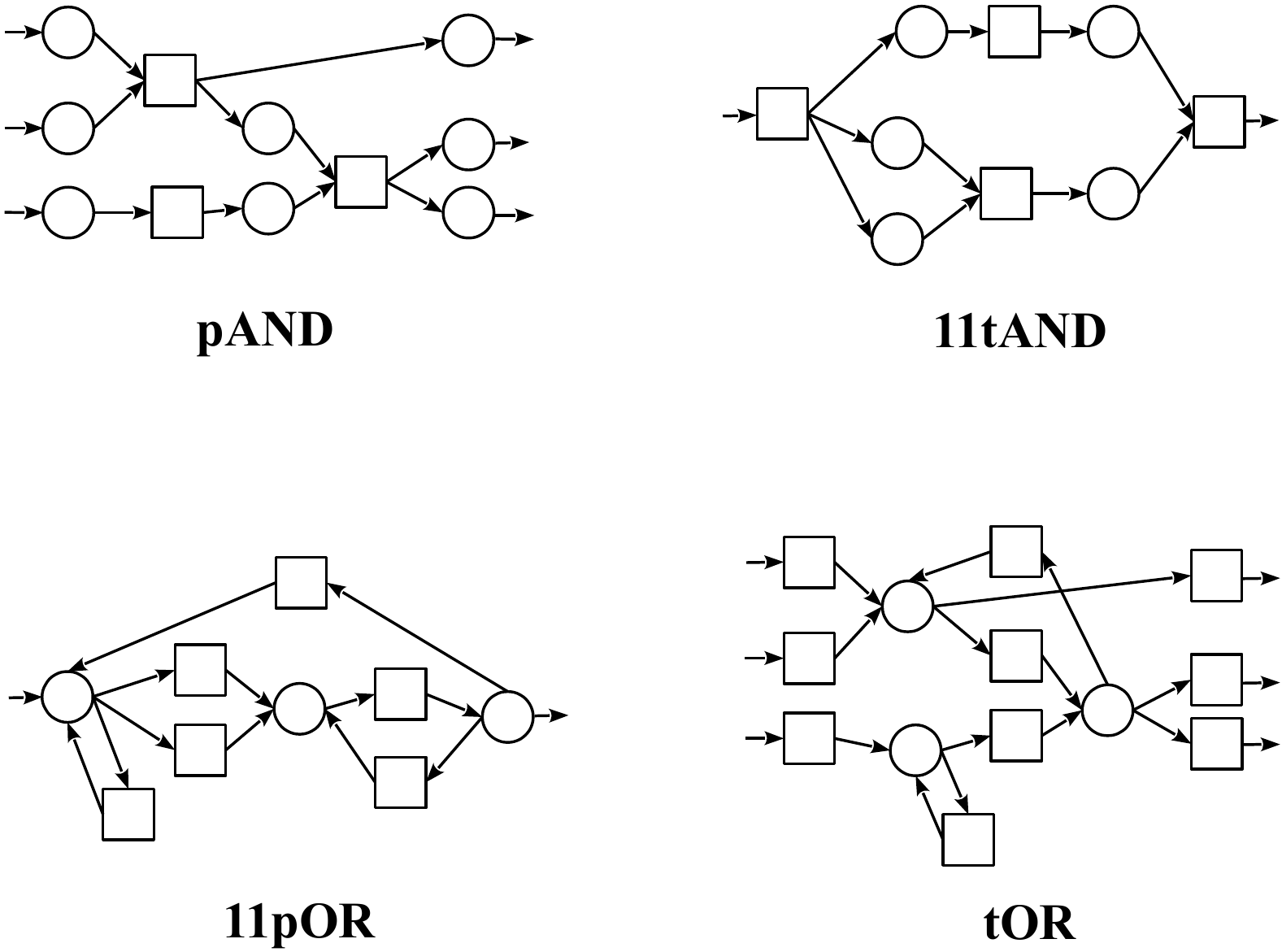}
\par\end{centering}
\caption{\label{fig:And-or-net11}Example nets from classes \textbf{pAND}, \textbf{11tAND}, \textbf{11pOR} and \textbf{tOR}}
\end{figure}
\par\end{center}

\begin{defn}[AND-OR net]
The class $\mathbf{S}(\mathbf{pAND}\cup\mathbf{11tAND}\cup\mathbf{11pOR}\cup\mathbf{tOR})$ we call the class of AND-OR nets.
\end{defn}

An example of the generation of an AND-OR net is shown in Figure~\ref{fig:And-or-net-example}, with the hierarchical decomposition in (a) and the resulting net in (b).

\begin{center}
\begin{figure}
\centering{}\includegraphics[width=0.35\textwidth]{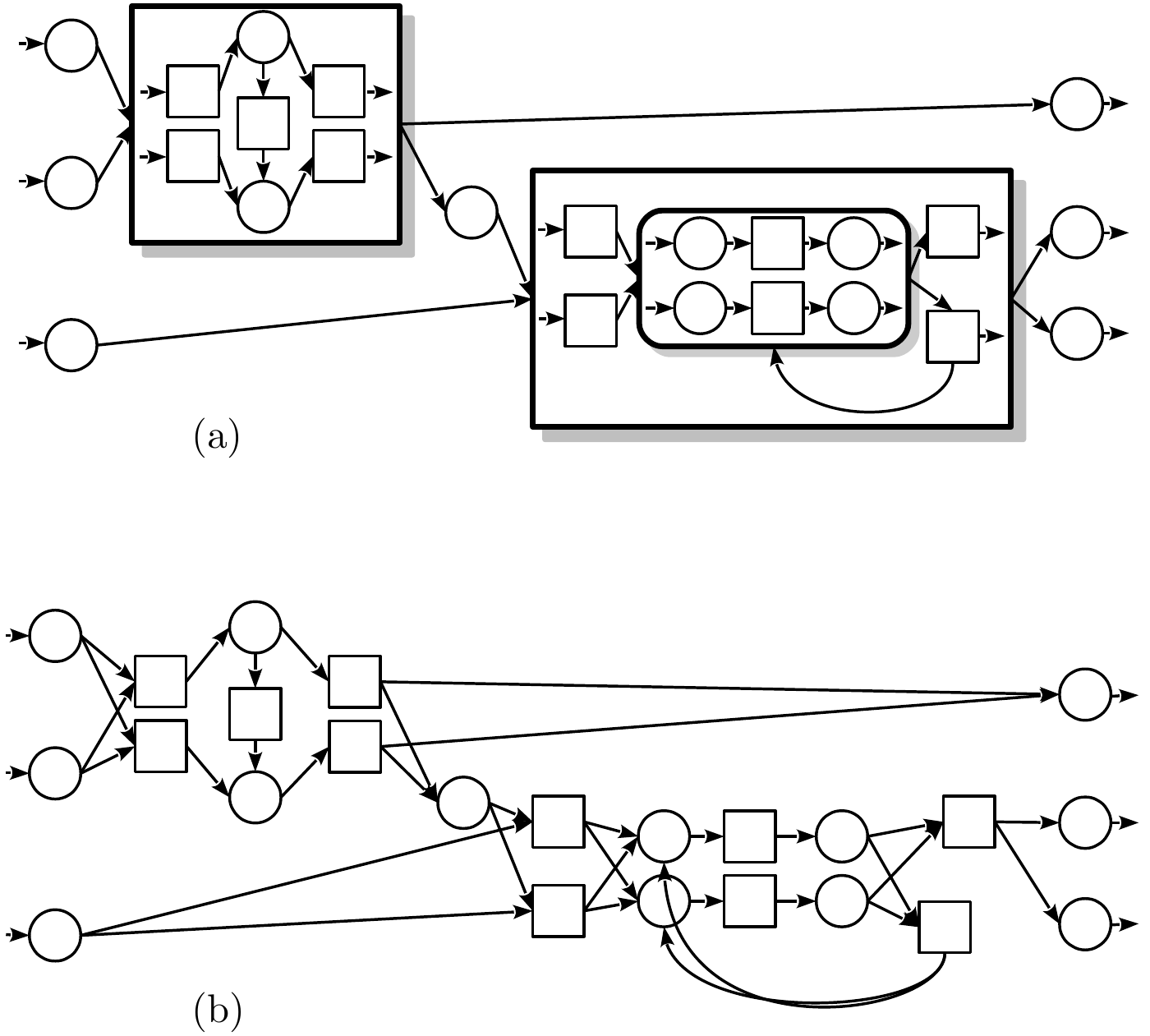}\caption{\label{fig:And-or-net-example}An example of the generation of an AND-OR net: (a) the hierarchical decomposition and (b) the resulting net}
\end{figure}
\par\end{center}

It can be shown that the one-input one-output tAND nets are not needed, i.e., we can remove them from the initial class without changing the set of nets that can be generated.
\begin{thm}
The tAND nets are redundant for generating AND-OR nets, i.e., $\mathbf{S}(\mathbf{pAND}\cup\mathbf{11tAND}\cup\mathbf{11pOR}\cup\mathbf{tOR})\mathbf{=S}(\mathbf{pAND}\cup\mathbf{11pOR}\cup\mathbf{tOR})$.\end{thm}
\begin{IEEEproof}
Recall that tAND nets do not contain cycles. Also note that if we take a one-input one-output tAND net with input transition $t_{i}$ and output transition $t_{o}$ and we remove the begin and end transition, then we are left with a pAND net with $I=t_{i}\bullet$ and $O=\bullet t_{o}$. So every one-input one-output tAND net can be generated by starting with an tOR net consisting of a transition followed by a place which is again followed by a transition, and then substituting the previously mentioned pAND net for the place in the middle.
\end{IEEEproof}

\begin{center}
\begin{figure}
\centering
\includegraphics[width=0.31\textwidth]{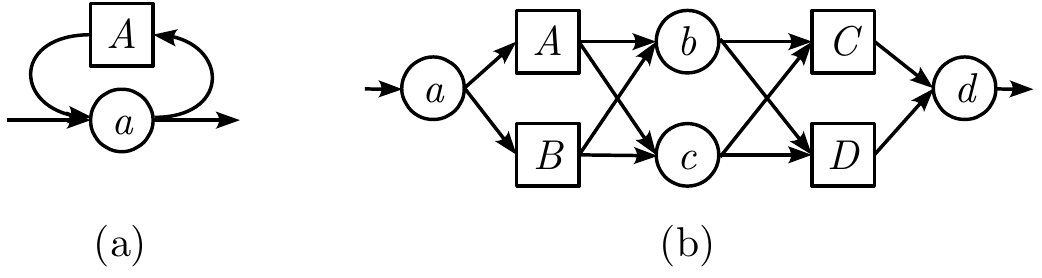}
\caption{\label{fig:counter-example-and-or-hyp}Examples showing the expressive power of certain classes}
\end{figure}
\end{center}

However, the one-input one-output pOR nets are not redundant, because a cycle containing the input and output nodes cannot be obtained in any other way.
\begin{thm}
The pOR nets are not redundant for generating all AND-OR nets, i.e., $\mathbf{S}(\mathbf{pAND\cup\mathbf{11tAND}}\cup\mathbf{11pOR}\cup\mathbf{tOR})\mathbf{\supsetneq S}(\mathbf{pAND\cup\mathbf{11tAND}}\cup\mathbf{tOR})$.\end{thm}
\begin{IEEEproof}
See the counterexample in Figure~\ref{fig:counter-example-and-or-hyp} (a). This one-input one-output pOR net cannot be generated by using pAND, one-input one-output tAND and tOR nets.
\end{IEEEproof}

Of course pAND nets and tOR nets are not redundant either, since they allow for multiple input and output nodes.

The AND-OR nets are very similar to the ST nets defined in~\cite{DBLP:conf/apn/HeeSV03} by van Hee et al. In fact, the class of ST nets is the strict subclass of $\mathbf{S}(\mathbf{11tAND}\cup\mathbf{11pOR})$ that disallows incoming edges for input nodes and outgoing edges for output nodes. It is clear that the class $\mathbf{S}(\mathbf{11tAND}\cup\mathbf{11pOR})$ is a proper subclass of the AND-OR nets since it only contains one-input one-output WF nets. However there are in addition also one-input one-output AND-OR nets that are not in $\mathbf{S}(\mathbf{11tAND}\cup\mathbf{11pOR})$ as is shown by the following theorem.

\begin{thm}
The class $\mathbf{S}(\mathbf{11tAND}\cup\mathbf{11pOR})$ does not contain all one-input one-output AND-OR nets.\end{thm}
\begin{IEEEproof}
The counterexample is given in Figure~\ref{fig:counter-example-and-or-hyp} (b). To show that it is an AND-OR net we consider its generation in reverse. The transitions $A$ and $B$ form an tOR net and can be contracted into a single transition. The same for the transitions $C$ and $D$. The places $b$ and $c$ form a pAND net and can be contracted into a single place. The result will be a linear net that is in fact both a pAND net and a one-input one-output pOR net. To see that the example net is not in $\mathbf{S}(\mathbf{11tAND}\cup\mathbf{11pOR})$ it can be verified that there is no proper subnet that is either in $\mathbf{11tAND}$ or $\mathbf{11pOR}$ and can be contracted into a single transition or place, respectively. \end{IEEEproof}

\section{Substitution soundness}\label{sec:subsoundness}

Recall that the purpose of this paper is to investigate the possibility to generate a large class of {*}-sound WF nets by using substitution. It is unfortunately not true that {*}-soundness is preserved by substitutions as defined in this paper. This is because of the possible outgoing edges of the output nodes. A counterexample is shown in Figure~\ref{fig:pc-tc-soundness} where the presented pWF net can be thought of as being constructed by substituting a {*}-sound net $N$, with input place $a$ and output place $c$, into an also {*}-sound sequential pWF net. As was discussed in the proof of Theorem~\ref{thm:tc-soundness} the resulting net is not 1-sound so also not {*}-sound. Therefore, we introduce a new notion of soundness called \emph{substitution soundness} and study its properties. As we will show in Section~\ref{sec:subsoundness_AND-OR} that all the basic classes of nets from the definition of AND-OR nets are substitution sound.

The intuition underlying substitution soundness is that it should not matter that during a run of a workflow net we remove seemingly ready tokens from output places. In other words, it should hold that if the net starts with $k$ tokens in the input places, reaches a marking with at least $k'\leq k$ tokens in each output place, and we remove these $k'$ tokens from each output place, then the net can still finish with $k-k'$ tokens in each output place.
\begin{defn}
[Substitution soundness] Let $N=(P,T,F,I,O)$ be a pWF net. We say that $N$ is \emph{substitution-sound} (or simply \emph{sub-sound}) if for all $k \geq k' \geq 0$ and every marking $m'$ it holds that if $k.I \stackrel{*}{\longrightarrow} (m'+k'.O)$ then $m' \stackrel{*}{\longrightarrow} (k-k').O$. We generalize this property to tWF nets and say that a tWF net $N$ is sub-sound if $\pc{N}$ is sub-sound. 
\end{defn}

We claim this is in some sense a necessary condition to construct 1-sound nets by substitution of nodes in 1-sound nets. In particular it can be shown that there is no weaker condition that is preserved by substitution and implies 1-soundness.

\begin{thm}
There is no property of pWF nets that (1) is strictly weaker then substitution soundness, i.e., it is implied by substitution soundness but not vice versa, (2) implies 1-soundness and (3) is preserved by substitution.
\end{thm}

\begin{IEEEproof}
Consider the class of pWF nets illustrated in Figure~\ref{fig:subsound-necessity} where a pWF net is defined for each value of $k$, which we will call $M_k$. Note that in $M_k$ the subnet defined by $b_i$ and $B_i$ are repeated $k$ times, and the same for the subnet defined by $D_i, E_i$ and $e_i$.

It can be easily observed that these nets are 1-sound, and in fact are substitution sound. Now consider a pWF net $N$ that is not substitution sound such that if we let it start with $k$ tokens in the input places and during its run remove $k' < k$ tokens from the output places then it cannot reach the final marking. If we substitute $N$ in $M_k$ for place $d$, i.e., we consider $M_k \otimes_d N$, then we obtain a net that is not 1-sound. To see this consider the following. We can let $M_k$ start with one token in $a$ and run until there are $k$ tokens in the input places of $N$. Then we can run $N$ until there are $k'$ tokens in its output places. These tokens can then be removed by firing $k$ times $E$. Since after this $N$ cannot reach a final state with $k - k'$ in its output places, it follows that the net  $M_k \otimes_d N$ cannot reach its final state.

The theorem now follows from the previous by the following reduction ad absurdum. Assume some property that satisfies (1), (2) and (3) at the same time. Observe that $M_k$ will satisfy this property since this property is weaker then substitution soundness. Also observe that there has to be pWF net $N$ that satisfies this property but is not substitution sound. By (3) it then follows that $M_k \otimes_d N$ also has the property an therefore by (2) that it is 1-sound. This, however, contradicts what we observed previously, namely that the result is not 1-sound.
\end{IEEEproof}

\begin{center}
\begin{figure}
\centering{}\includegraphics[width=0.45\textwidth]{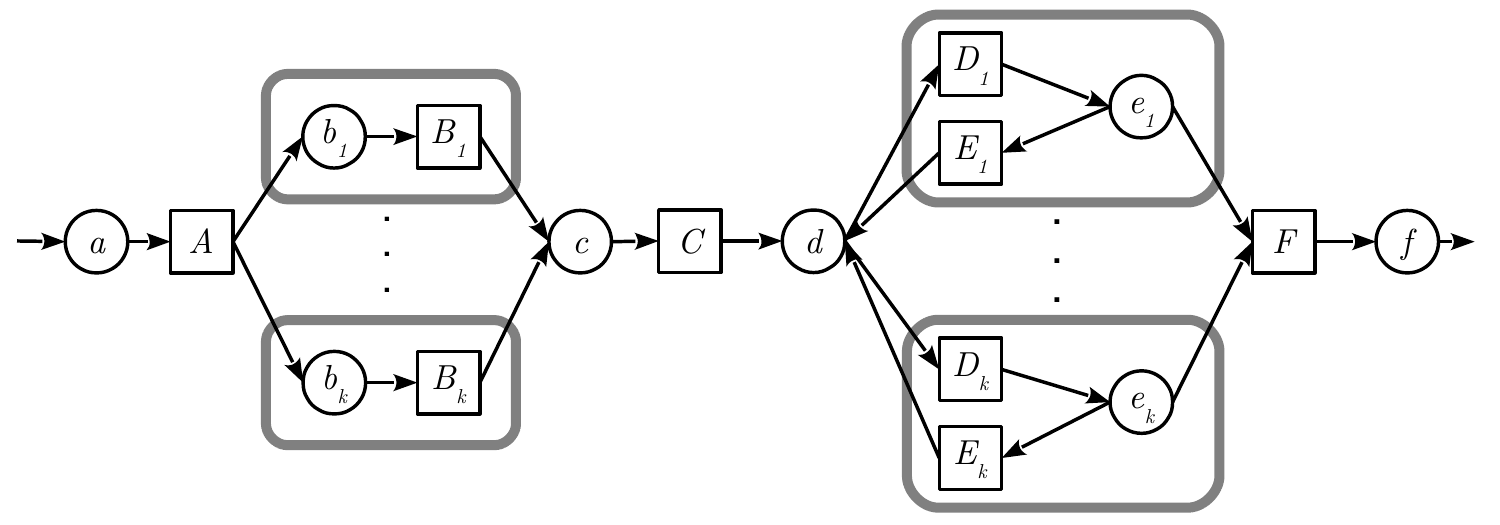}
\caption{\label{fig:subsound-necessity}Illustration of necessity of substitution soundness}
\end{figure}
\par\end{center}

Observe that the previous theorem does not establish that substitution soundness is necessary in the sense that every property that satisfies the three conditions is stronger then substitution soundness. This is however the case if we add the requirement that the property must hold for the nets in Figure~\ref{fig:subsound-necessity}, i.e., these nets should be in the class of nets that we intend to generate by substitution, which seems a reasonable requirement. 

Now we prove that sub-soundness is sufficient for constructing {*}-sound nets by substitution. First, note that the case where $k'=0$ describes {*}-soundness and so sub-soundness implies {*}-soundness. Furthermore, on many classes of nets the two notions of soundness coincide, as
is shown by the following two lemmas.
\begin{lem}
\label{lem:sub-is-star-if-no-out}For every pWF net $N$ such that all output places have no outgoing edges it holds that $N$ is {*}-sound iff $N$ is sub-sound.\end{lem}
\begin{IEEEproof}
As already argued it holds that sub-soundness implies {*}-soundness, so the converse remains to be shown. Let $N=(P,T,F,I,O)$. Assume that $k.I\stackrel{*}{\longrightarrow}(m+k'.O)$ for some $k'$ such that $k\geq k'\geq0$. By {*}-soundness it holds for some $\sigma$ that $(m+k'.O)\stackrel{\sigma}{\longrightarrow}k.O$. However, since the places in $O$ have no outgoing edges none of the transitions in $\sigma$ consumes any of their tokens and so $m\stackrel{\sigma}{\longrightarrow}(k-k').O$. 
\end{IEEEproof}
Note that the restriction mentioned in Lemma~\ref{lem:sub-is-star-if-no-out} is included in the classical definition of WF net by van der Aalst~\cite{Aalst1998workflow}. However, with this restriction we would not be able to generate all AND-OR nets, not even all those that satisfy this restriction. In particular we would not be able to do arbitrary loop additions. As an example consider Figure~\ref{fig:counter-example-and-or-hyp} (b) where we would not be able to add a loop to place $b$. Note that a similar restriction is not necessary for tWF nets because for them the soundness properties are defined by place completion. Recall also that for tOR nets the output transitions cannot have outgoing edges by definition and for one-input one-output tAND nets this follows from the facts that AND nets are acyclic and that in a tWF nets it is possible to reach one of the output transitions from every place and transition.

\begin{lem}
\label{lem:tWF-subsound-is-starsound}For every tWF net N it holds that N is {*}-sound iff N is sub-sound.\end{lem}
\begin{IEEEproof}
As already argued, it is enough to show that {*}-soundness implies sub-soundness. A tWF net $N$ is by definition sub-sound iff $\pc{N}$ is sub-sound. Since in $\pc{N}$ the output place has no outgoing edges it follows from Lemma~\ref{lem:sub-is-star-if-no-out} that $\pc{N}$ is sub-sound iff it is {*}-sound. Finally, by definition it holds that $\pc{N}$ is {*}-sound iff $N$ is {*}-sound.
\end{IEEEproof}

We now proceed with showing that sub-soundness is sufficient for constructing *-sound nets by substitution. In Theorems~\ref{thm:place-subst-soundness-in-pWF} and~\ref{thm:place-subst-soundness-in-tWF} we show that sub-soundness is preserved while substituting places in pWF nets. By this we mean that if we take a sub-sound pWF net or tWF net and substitute a place in it by another sub-sound pWF net, we again obtain a sub-sound pWF net or tWF net, respectively.  Similarly, in Theorems~\ref{thm:trans-subst-soundness-in-pWF} and~\ref{thm:trans-subst-soundness-in-tWF} we show that sub-soundness is also preserved while substituting transitions in pWF nets and tWF nets respectively, i.e., if we take a sub-sound pWF net or tWF net and substitute a transition in it by another sub-sound tWF net, we again obtain a sub-sound pWF net or tWF net, respectively.

\begin{thm}
\label{thm:place-subst-soundness-in-pWF}If a pWF net $N=(P_{N},T_{N},F_{N},I_{N},O_{N})$ and a disjoint pWF net $M=(P_{M},T_{M},F_{M},I_{M},O_{M})$ are sub-sound, then for any $p\in P_{N}$ the net $N\otimes_{p}M$ is also sub-sound.
\end{thm}
\begin{IEEEproof}
Let $N_{NM}=N\otimes_{p}M=(P_{NM},T_{NM},F_{NM},I_{NM},O_{NM})$. We define $\mathbf{S}(M,k)$ as the set of markings $m_M$ of $M$ that represent the fact that there are still $k$ ``threads'' active in $M$ after possibly having started with more threads but some of them ended by the removal of tokens from $O'$, i.e., for some $k'\geq k$ it holds that $k'.I'\stackrel{*}{\longrightarrow}_{M}m_{M}+(k'-k).O_{M}$. We define a simulation relation $\sim\subseteq\mathbf{M}_{N}\times\mathbf{M}_{NM}$ such that $m_{N}{\sim}m_{NM}$ represents the fact that $m_{N}$ is the same as $m_{NM}$ except that all (say $k$) tokens are removed from $p$ and replaced by some marking from $\mbox{\textbf{S}}(M,k)$, i.e., $m_{NM}=m_{N}-[p^{k}]+m_{M}^{k}$ for some $m_{M}^{k}\in\mbox{\textbf{S}}(M,k)$ with $k=m_{N}(p)$.

We first discuss the idea of the proof and then follow with the laborious details.

\begin{figure*}
\centering{}\includegraphics[width=1\textwidth]{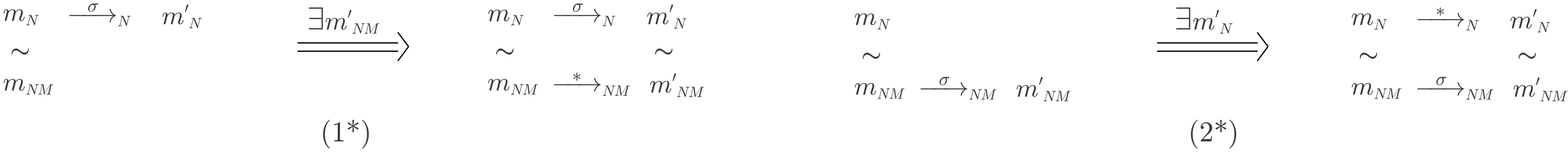}\caption{\label{fig:proof_idea1}$\sim$ indeed defines a kind of bisimilarity}
\end{figure*}

It can be shown that $\sim$ indeed defines a kind of bisimilarity, i.e., (see Figure~\ref{fig:proof_idea1}) it holds that:
\begin{description}
\item[(B*):] if $m_{N}\stackrel{\sigma}{\longrightarrow}_{N}m'_{N}$ and $m_{N}{\sim}m_{NM}$, then there is a marking $m'_{NM} \in \mathbf{M_{NM}}$ such that $m_{NM}\stackrel{*}{\longrightarrow}_{NM}m'_{NM}$ and $m'_{N}{\sim}m'_{NM}$ and
\item[(C*):] if $m_{NM}\stackrel{\sigma}{\longrightarrow}_{NM}m'_{NM}$ and $m_{N}{\sim}m_{NM},$ then there is a marking $m'_{N} \in \mathbf{M_{N}}$ such that $m_{N}\stackrel{*}{\longrightarrow}_{N}m'_{N}$ and $m'_{N}{\sim}m'_{NM}$.
\end{description}
This can be shown with induction on the length of $\sigma$ where for each transition $t$ in $\sigma$ we distinguish for (B*) the cases where $p\in\bullet_{N}t$ or not and $p\in t\bullet_{N}$ or not, and for (C*) we distinguish the cases where $t$ is a transition in $N$ or $M$. 

We then can show the sub-soundness of $N\otimes_{p}M$ using (B*) and (C*). The idea of this part is as follows (see Figure~
\ref{fig:proof_idea2}). Assume that $k.I_{NM}\stackrel{*}{\longrightarrow}_{NM}(m_{NM}+k'.O_{NM})$ with $k\geq k'\geq0$. By (C*) and the fact that $k.I_{N} {\sim}k.I_{NM}$ it then follows that $k.I_{N}\stackrel{*}{\longrightarrow}_{N}m_{N}$ such that $m_{N}\sim(m_{NM}+k'.O_{NM})$. We can show that we can assume that $m_{N}=m'_{N}+k'.O_{N}$ with $m'_{N}$ a marking of $N$. By the sub-soundness of $N$ it holds that $m'_{N}\stackrel{*}{\longrightarrow}_{N}(k-k').O_{N}$. At the same time by the definition of $\sim$ it follows that $m'_{N}{\sim}m_{NM}$. Using the last two from (B*) it then follows that $m_{NM}\stackrel{*}{\longrightarrow}_{NM}m'_{NM}$ such that $(k-k').O_{N}{\sim}m'_{NM}$. Although similar to $(k-k').O_{N}$, the $m'_{NM}$ does not have to be the final marking $(k-k').O_{NM}$, yet by using sub-soundness of $M$ it can be shown that $m'_{NM}\stackrel{*}{\longrightarrow}_{NM}(k-k').O_{NM}$.

We proceed with the proof of (B*) and (C*). We first show that
\begin{description}
\item[(A):] if $m_{N}{\sim}m_{NM}$, $m_{N}\stackrel{t}{\longrightarrow}_{N}m'_{N}$ and $m_{NM}\stackrel{t} {\longrightarrow}_{NM}m'_{NM}$, then $m'_{N}{\sim}m'_{NM}$.
\end{description}
We then use (A) to show
\begin{description}
\item[(B):] if $m_{N}{\sim}m_{NM}$ and $m_{N}\stackrel{t}{\longrightarrow}_{N}m'_{N}$, then there is a marking $m'_{NM}$ such that $m_{NM}\stackrel{*}{\longrightarrow}_{NM}m'_{NM}$ and $m_{NM}{\sim}m'_{NM}$, and
\item[(C):] if $m_{N}{\sim}m_{NM}$ and $m_{NM}\stackrel{t}{\longrightarrow}_{NM}m'_{NM}$, then there is a marking $m'_{N}$ such that $m_{N}\stackrel{*}{\longrightarrow}_{N}m'_{N}$ and $m'_{N}{\sim}m'_{NM}$. 
\end{description}
Then with induction we generalize (B) and (C) to (B*) and (C*), respectively. We now proceed with the proofs of claim (A) , (B) and (C).

\textbf{Proof of claim (A):} Assume that $m_{N}\stackrel{t}{\longrightarrow}_{N}m'_{N}$ and $m_{NM}\stackrel{t}{\longrightarrow}_{NM}m'_{NM}$. We also assume $m_{N}{\sim}m_{NM}$, which by definition gives $m_{NM}=m_{N}-[p^{k}]+m_{M}^{k}$ for some $m_{M}^{k}\in\mbox{\textbf{S}}(M,k)$ with $k=m_{N}(p)$. After firing $t$ in $m_{NM}$ we get $m'_{NM}=(m_{N}-[p^{k}]+m_{M}^{k}-\bullet_{NM}t+t\bullet_{NM})$. We consider the four cases for whether $p\in\bullet_{N}t$ or not, and $p\in t\bullet_{N}$ or not:

(i) Assume $p \notin \bullet_{N}t$ and $p\notin t\bullet_{N}$. In that case $\bullet_{NM}t=\bullet_{N}t$ and $t\bullet_{NM}=t\bullet_{N}$ and therefore $m'_{NM}=(m_{N}-\bullet_{N}t+t\bullet_{N}-[p^{k}]+m_{M}^{k})$ and since $m_{N}\stackrel{t}{\longrightarrow}_{N}m'_{N}$ it follows that $m'_{NM}=(m'_{N}-[p^{k}]+m_{M}^{k})$. Now, it remains to be shown that $m'_{N}(p)=k$ which follows from $m_{N}(p)=k$ and $p\notin \bullet_{N}t$ and $p\notin t\bullet_{N}$. This concludes that $m'_{N}{\sim}m'_{NM}$. 

(ii) Assume $p\in\bullet_{N}t$ and $p\notin t\bullet_{N}$. In that case $t\bullet_{NM}=t\bullet_{N}$ and from the construction of the substitution it follows that $\bullet_{NM}t=\bullet_{N}t-[p]+O_{M}$ and therefore $m'_{NM}=(m_{N}-\bullet_{N}t+t\bullet_{N}-[p^{(k-1)}]+m_{M}^{k}-O_{M})$ and since $m_{N}\stackrel{t}{\longrightarrow}_{N}m'_{N}$ it follows that $m'_{NM}=(m'_{N}-[p^{(k-1)}]+m_{M}^{k}-O_{M})$. Then, it holds that (a) $m'_{N}(p)=k-1$ because $m_{N}(p)=k$ and $p\in\bullet_{N}t$ and $p\notin t\bullet_{N}$, and (b) $m_{M}^{k}-O_M\in\mathbf{S}(M,k-1)$ since $m_{M}^{k}\in\mathbf{S}(M,k)$. Observe that $m_{M}^{k}-O_M$ is a valid state, i.e., there is a non-negative number of tokens in each place, because we assumed $t$ is enabled in $m_{NM}$ as well $p \in \bullet_N t$ and the $m_{M}^{k}$ component of $m_{NM}$ covers tokens in places from $P_M$. From (a) and (b) it follows that $m'_{N}{\sim}m'_{NM}$.

(iii) Assume $p\notin \bullet_{N}t$ and $p\in t\bullet_{N}$. In that case $\bullet_{NM}t=\bullet_{N}t$ and from the construction of the substitution $t\bullet_{NM}=t\bullet_{N}-[p]+I_{M}$ and therefore $m'_{NM}=(m_{N}-\bullet_{N}t+t\bullet_{N}-[p^{(k+1)}]+m_{M}^{k}+I_{M})$ and since $m_{N}\stackrel{t}{\longrightarrow}_{N}m'_{N}$ it follows that $m'_{NM}=(m'_{N}-[p^{(k+1)}]+m_{M}^{k}+I_{M})$. Then, it holds that (a) $m'_{N}(p)=k+1$ because $m_{N}(p)=k$ and $p\notin \bullet_{N}t$ and $p\in t\bullet_{N}$, and (b) $m_{M}^{k}+I_{M}\in\mathbf{S}(M,k+1)$ since $m_{M}^{k}\in\mathbf{S}(M,k)$. From (a) and (b) it follows that $m'_{N}{\sim}m'_{NM}$.

(iv) Assume $p\in\bullet_{N}t$ and $p\in t\bullet_{N}$. In that case $\bullet_{NM}t=\bullet_{N}t-[p]+O_{M}$ and $t\bullet_{NM}=t\bullet_{N}-[p]+I_{M}$ and therefore $m'_{NM}=(m_{N}-\bullet_{N}t+t\bullet_{N}-[p^{k}]+m_{M}^{k}-O_{M}+I_{M})$ and since $m_{N}\stackrel{t}{\longrightarrow}_{N}m'_{N}$ it follows that $m'_{NM}=(m'_{N}-[p^{k}]+m_{M}^{k}-O_{M}+I_{M})$. Then, it holds that (a) $m'_{N}(p)=k$ because $m_{N}(p)=k$ and $p\in\bullet_{N}t$ and $p\in t\bullet_{N}$, and (b) $m_{M}^{k}-O_{M}+I_{M}\in\mathbf{S}(M,k)$ since $m_{M}^{k}\in\mathbf{S}(M,k)$. Observe that $m_{M}^{k}-O_{M}+I_{M}$ is a valid state for the same reasons as in (ii). From (a) and (b)
it follows that $m'_{N}{\sim}m'_{NM}$.

We have now covered all possible cases and in each of them concluded that $m'_{N}{\sim}m'_{NM}$ which finishes the proof of (A).

\textbf{Proof of claim (B):} Assume that $m_{N}\stackrel{t}{\longrightarrow}_{N}m'_{N}$, which by definition gives $\bullet_{N}t\leq m_{N}$. We also assume $m_{N}{\sim}m_{NM}$, which by definition gives $m_{NM}=m_{N}-[p^{k}]+m_{M}^{k}$ for some $m_{M}^{k}\in\mbox{\textbf{S}}(M,k)$ with $k=m_{N}(p)$. Since $m_{M}^{k}\in\mbox{\textbf{S}}(M,k)$ and $M$ is sub-sound, it holds that $m_{M}^{k}\stackrel{*}{\longrightarrow}_{M}k.O_{M}$, and since $M$ is embedded in $NM$, it follows that $m_{NM}\stackrel{*}{\longrightarrow}_{NM}m_{N}-[p^{k}]+k.O_{M}$. Note also that $m_{NM}{\sim}m_{N}-[p^{k}]+k.O_{M}$, because by definition $k.O_{M}\in\mbox{\textbf{S}}(M,k)$. Now we observe that since $t$ is a transition in $N$, if $t$ is enabled in $m_{N}$ for $N$, by construction of $NM$ it is also enabled in $m_{N}-[p^{k}]+k.O_{M}$ for $NM$ regardless of $p\in\bullet_{N}t$. It follows that there is a marking $m'_{NM}$  such that $m_{N}-[p^{k}]+k.O_{M}\stackrel{t}{\longrightarrow}_{NM}m'_{NM}$ and thus $m_{NM}\stackrel{*}{\longrightarrow}_{NM}m'_{NM}$. By (A) it follows that $m'_{N}{\sim}m'_{NM}$, which concludes the proof of (B).

\textbf{Proof of claim (C):} Assume that $m_{NM}\stackrel{t}{\longrightarrow}_{NM}m'_{NM}$. We also assume $m_{N}{\sim}m_{NM}$, which by definition gives $m_{NM}=m_{N}-[p^{k}]+m_{M}^{k}$ for some $m_{M}^{k}\in\mbox{\textbf{S}}(M,k)$ with $k=m_{N}(p)$. We consider the two possible cases: $t$ is a transition in $N$, and $t$ is a transition in $M$.

(i) Assume that $t$ is a transition in $N$. Since $t$ was enabled in $m_{NM}$ for $NM$, i.e., $\bullet_{NM}t\leq m_{NM}$, it will also be enabled in $m_{N}$ for $N$, i.e., $\bullet_{N}t\leq m_{N}$. This can be shown as follows. Suppose $p\notin \bullet_{N}t$, then $\bullet_{NM}t=\bullet_{N}t$. Since $m_{M}^{k}$ contains only places in $M$ it follows from $\bullet_{NM}t\leq m_{NM}=m_{N}-[p^{k}]+m_{M}^{k}$ that $\bullet_{N}t=\bullet_{NM}t\leq m_{N}$. Suppose on the other hand that $p\in\bullet_{N}t$, then $\bullet_{NM}t=\bullet_{N}t-[p]+O_{M}$ and we get $\bullet_{N}t-[p]+O_{M}\leq m_{N}-[p^{k}]+m_{M}^{k}$. Both sides of this inequality can be limited to $N$ by omitting components not from $P_N$, giving $\bullet_{N}t - [p] \leq m_{N} - [p^k]$. Since in this case $k \geq 1$ we get $\bullet_N t \leq m_N$. Now, since $t$ is enabled in $m_{N}$ for $N$ there will be a marking $m'_{N}$ such that $m_{N}\stackrel{t}{\longrightarrow}_{N}m'_{N}$ and it follows by (A) that $m'_{N}{\sim}m'_{NM}$.

(ii) Assume that $t$ is a transition in $M$. In this case the marking we are looking for is $m_N$ itself. Since $\bullet_{NM}t$ are all places in $M$, it follows that $t$ is enabled in $m_{M}^{k}$ for $M$. So there is $m'_{M}$ such that $m_{M}^{k}\stackrel{t}{\longrightarrow}_{M}m'_{M}$ and because $t\bullet_{NM}$ are also all places in $M$, we have $m'_{M}=m_{M}^{k}-\bullet_{M}t+t\bullet_{M}$. Now from our assumptions it follows that $m'_{NM}=m_{NM}-\bullet_{M}t+t\bullet_{M}=m_{N}-[p^{k}]+m_{M}^{k}-\bullet_{M}t+t\bullet_{M}=m_{N}-[p^{k}]+m'_{M}$. Since $m_{M}^{k}\in\mbox{\textbf{S}}(M,k)$ and $m_{M}^{k}\stackrel{t} {\longrightarrow}_{M}m'_{M}$, then it also holds that $m'_{M}\in\mbox{\textbf{S}}(M,k)$. From the assumption that $k=m_{N}(p)$, it follows that $m_{N}{\sim}m'_{NM}$, and obviously it also holds that $m_{N}\stackrel{*}{\longrightarrow}_{N}m_{N}$.

Since in both possible cases it follows that there is a marking $m'_{N}$ such that $m_{N}\stackrel{*}{\longrightarrow}_{N}m'_{N}$ and $m'_{N}{\sim}m'_{NM}$, we can conclude that this always follows, which concludes the proof of (C).

\textbf{Proof of claim (B*) and (C*):} We can straightforwardly generalize (B) and (C) by using induction on the length of $\sigma$ and show  that (1{*}) if $m_{N}\stackrel{\sigma}{\longrightarrow}_{N}m'_{N}$ and $m_{N}{\sim}m{}_{NM}$ then there is a marking $m'_{NM}$ such that $m{}_{NM}\stackrel{*}{\longrightarrow}_{NM}m'_{NM}$ and $m'_{N}{\sim}m'_{NM}$ and (2{*}) if $m{}_{NM}\stackrel{\sigma}{\longrightarrow}_{NM}m'_{NM}$ and $m_{N}{\sim}m{}_{NM}$, then there is a marking $m'_{N}$ such that $m{}_{N}\stackrel{*}{\longrightarrow}_{N}m'_{N}$ and $m'_{N}{\sim}m'_{NM}$.

\begin{figure*}
\centering{}\includegraphics[width=1\textwidth]{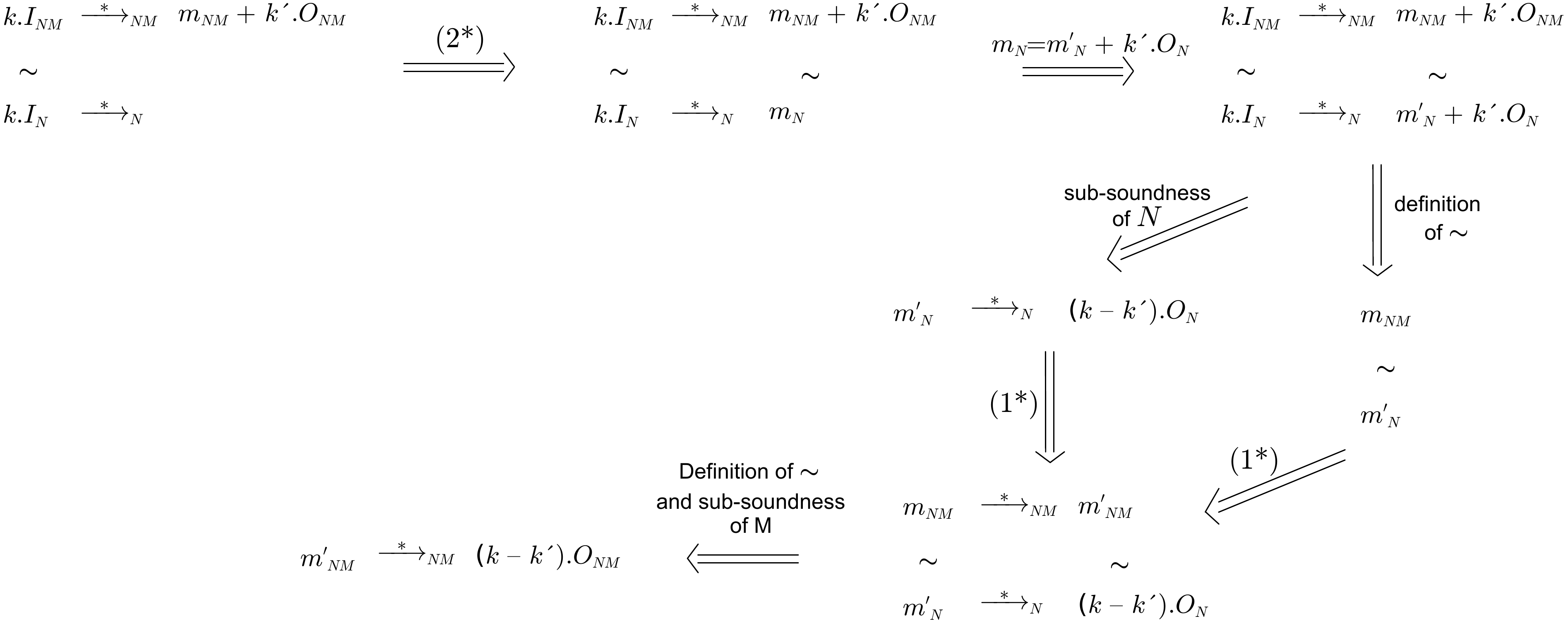}\caption{\label{fig:proof_idea2}Structure of the proof of Theorem~\ref{sec:subsoundness}}
\end{figure*}

We proceed with the proof of the final part, i.e., prove the sub-soundness of $N\otimes_{p}M$ using (B*) and (C*).

\textbf{Proof of sub-soundness of  $N\otimes_{p}M$:} The full structure of the reasoning is presented in Figure~\ref{fig:proof_idea2}. Assume that $k.I_{NM}\stackrel{*}{\longrightarrow}_{NM}(m_{NM}+k'.O_{NM})$ with $k\geq k'\geq0$. Since $I_{NM}=I_{N}$ if $p\notin I_{N}$ and $I_{NM}=I_{N}-[p]+I_{M}$ if $p\in I_{N}$, it holds that $k.I_{N}{\sim}k.I_{NM}$. By (C*) it then follows that $k.I_{N}\stackrel{*}{\longrightarrow}_{N}m_{N}$ such that $m_{N}\sim(m_{NM}+k'.O_{NM})$. 

We now construct $m'_{N}=m_{N}-k'.O_{N}$ and show that $m'_{N}{\sim}m_{NM}$ regardless of $p\notin O_{N}$ or $p\in O_{N}$. We start with showing the fact that $m'_{N}$ is a valid state, i.e., $m_{N}$ includes the tokens we are subtracting from it. Since $m_{N}\sim(m_{NM}+k'.O_{NM})$ for $k''=m_{N}(p)$ there is $m_{M}^{k''}\in\mathbf{S}(M,k'')$ such that $m_{NM}+k'.O_{NM}=m_{N}-[p^{k''}]+m_{M}^{k''}$. This gives $m_{N}=m_{NM}+k'.O_{NM}+[p^{k''}]-m_{M}^{k''}$. Let us consider two cases. For $p\notin O_{N}$, in which case $O_{NM}=O_{N}$, this gives $m_{N}=m_{NM}+k'.O_{N}+[p^{k''}]-m_{M}^{k''}$. It remains to observe that substracting the $m_{M}^{k''}$ component does not remove any tokens from $O_{N}$ because from disjointness of $N$ and $M$ we have $O_{N}\cap P_{M}=\emptyset$. For $p\in O_{N}$, in which case $O_{NM}=O_{N}-[p]+O_{M}$, we get $m_{N}=m_{NM}+k'.O_{N}+[p^{k''}]-[p^{k'}]+k'.O_{M}-m_{M}^{k''}$. Both sides of the equality have to include the same number of tokens in $p$. Since $m_{NM}$ marks only places from $P_{NM}=(P_{N}\setminus\{p\})\cup P_{M}$ and $k'.O_{M}-m_{M}^{k''}$ only places from $P_{M}$ (and $p\notin P_{M}$), all the tokens in $p$ are given by $k'.O_{N}+[p^{k''}]-[p^{k'}]$. It remains to show that $k''\geq k'$. This follows from further examination of the equality $m_{N}=m_{NM}+k'.O_{N}+[p^{k''}]-[p^{k'}]+k'.O_{M}-m_{M}^{k''}$. This time we look at the number of tokens in $O_{M}$. On the left-hand side there are clearly none. On the right hand side there are $k'$ introduced by $k'.O_{M}$, and the only negative component $m_{M}^{k''}$ substract no more than $k''$ of such tokens.

Now we continue with showing that $m'_{N}{\sim}m_{NM}$. This time from $m_{NM}+k'.O_{NM}=m_{N}-[p^{k''}]+m_{M}^{k''}$ we conclude $m_{NM}=m_{N}-[p^{k''}]+m_{M}^{k''}-k'.O_{NM}$ and again consider the two cases for $p\notin O_{N}$ or $p\in O_{N}$. If $p\notin O_{N}$, then $O_{NM}=O_{N}$ and so $m_{NM}=m_{N}-k'.O_{N}-[p^{k''}]+m_{M}^{k''}=m'_{N}-[p^{k''}]+m_{M}^{k''}$ and $m'_N(p)=m_N(p)-k'.O_N(p)=k''-0=k''$ so $m'_{N}{\sim}m_{NM}$. If $p\in O_{N}$, then $O_{NM}=O_{N}-[p]+O_{M}$ and so $m_{NM}=m_{N}-[p^{k''}]+m_{M}^{k''}-k'.O_{NM}=m_{N}-[p^{k''}]+m_{M}^{k''}-k'.O_{N}+[p^{k'}]-k'.O_{M}=m_{N}-k'.O_{N}-[p^{k''-k'}]+m_{M}^{k''}-k'.O_{M}=m'_{N}-[p^{k''-k'}]+m_{M}^{k''}-k'.O_{M}$, so also then we can conclude that $m'_{N}{\sim}m_{NM}$ because $m_{M}^{k''}-k'.O_{M}\in\mathbf{S}(M,k''-k')$ and $k''-k'=(m_{N}-k'.O_{N})(p)=m'_{N}(p)$.

By the sub-soundness of $N$ it then holds that $m'_{N}\stackrel{*}{\longrightarrow}_{N}(k-k').O_{N}$. From (B*) it follows that $m_{NM}\stackrel{*}{\longrightarrow}_{NM}m'_{NM}$ such that $(k-k').O_{N}{\sim}m'_{NM}$, that is $m'_{NM}=(k-k').O_{N}-[p^{x}]+m_{M}^{x}$ with $m_{M}^{x}\in\mathbf{S}(M,x)$ and $x=(k-k').O_{N}(p)$. If $p\notin O_{N}$, then $x=0$ and $O_{N}=O_{NM}$, and therefore $m'_{NM}=(k-k').O_{NM}$. If $p\in O_{N}$, then $x=k-k'$ and therefore $m'_{NM}=(k-k').O_{N}-[p^{k-k'}]+m_{M}^{k-k'}$. Because $M$ is sub-sound, it holds that $m_{M}^{k-k'}\stackrel{*}{\longrightarrow}_{M}(k-k').O_{M}$, and since $M$ is embedded in $N$ and in this case $O_{NM}=O_{N}-[p]+O_{M}$, it follows that $m'_{NM} \stackrel{*}{\longrightarrow}_{NM} (k-k').O_{N}-[p^{k-k'}]+(k-k').O_{M}=(k-k').O_{NM}$. This way we have shown that in all cases $m_{NM} \stackrel{*}{\longrightarrow}_{NM} m'_{NM} \stackrel{*}{\longrightarrow}_{NM} (k-k').O_{NM}$ which concludes the proof.
\end{IEEEproof}

We now proceed with the case for place substitution in tWF nets. For that we will use the following lemma.

\begin{figure*}
\centering{}\includegraphics[width=0.50\textwidth]{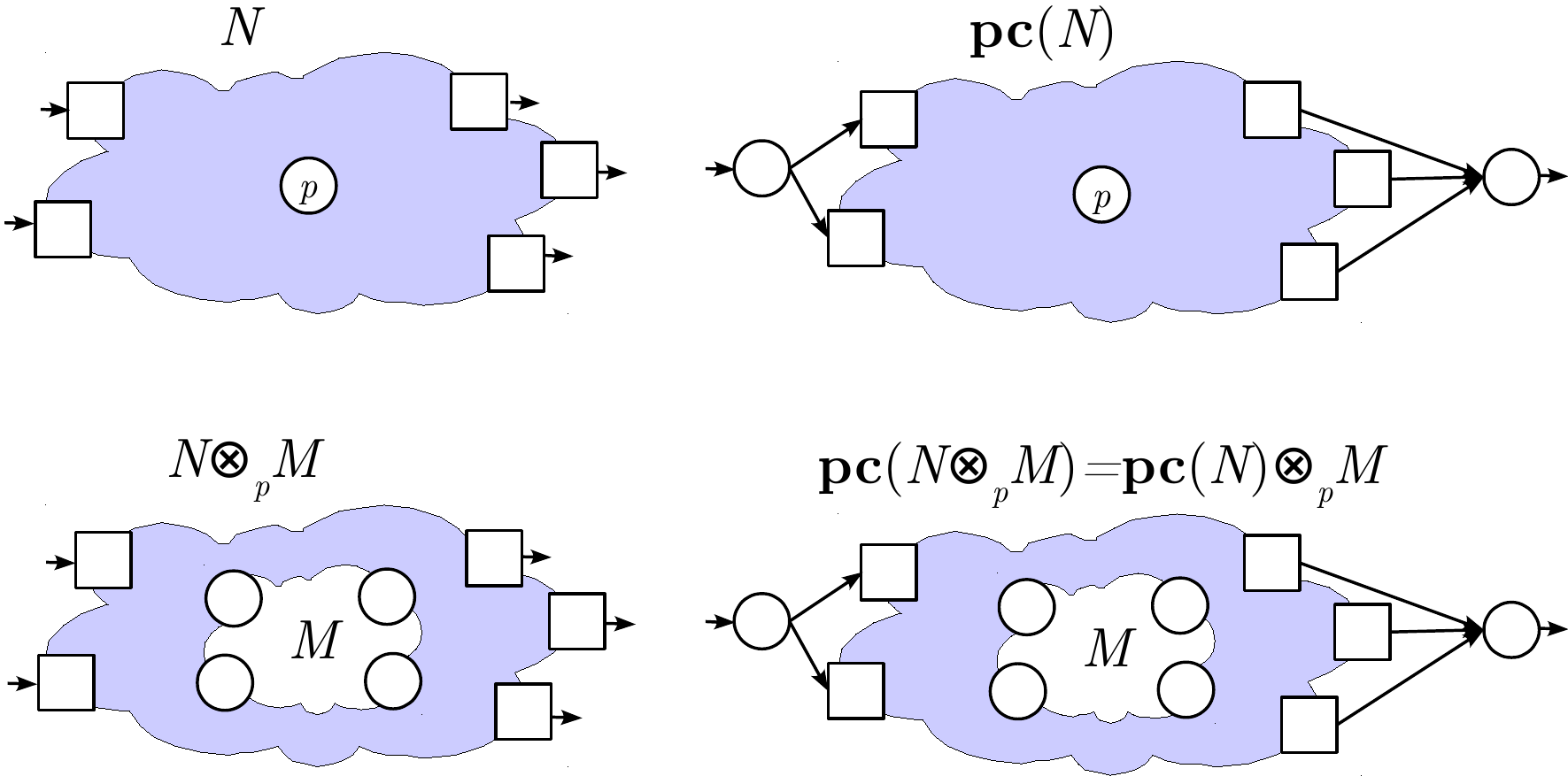}\caption{\label{fig:place-compl-of-place-subst}Place completion is semi-distributive in respect to place substitution}
\end{figure*}

\begin{lem}
\label{lem:commut-pc-subst}For every tWF net $N$ with a place $p$ and a disjoint pWF net $M$ it holds that $\pc{N\otimes_{p}M}=\pc{N}\otimes_{p}M$. \end{lem}
\begin{IEEEproof}
Let $N=(P_{N},T_{N},F_{N},I_{N},O_{N})$ with $p\in P_{N}$ and $M=(P_{M},T_{M},F_{M},I_{M},O_{M})$. In both cases the same nodes are added, viz., those of $M$ and $p_{i}$ and $p_{o}$, see Figure~\ref{fig:place-compl-of-place-subst}. Clearly the edges $F_{M}$ are added in the same way. Also in both cases afterward $p_{i}\bullet=I_{N}$ and $\bullet p_{o}=O_{N}$ because $N$ is a tWF net and $p\notin I_{N}$ and $p\notin O_{N}$. For nodes $p'\in I_{M}$ it holds in both cases that afterward $\bullet p'=\bullet_{N}p$ if $p\notin I_{N}$ and $\bullet p'=\{p_{i}\}$ if otherwise. Similarly for nodes $p'\in O_{M}$ afterward $p'\bullet=p\bullet_{N}$ if $p\notin O_{N}$ and $p'\bullet=\{p_{o}\}$. Finally, in both cases the final input set is $\{p_{i}\}$ and the final output set is $\{p_{o}\}$.\end{IEEEproof}

With this lemma we are able to show that sub-soundness is preserved while substituting place in tWF nets by converting it to the already proven case for pWF nets (see Theorem~\ref{thm:place-subst-soundness-in-pWF}).

\begin{thm}
\label{thm:place-subst-soundness-in-tWF}If a tWF net $N$ is sub-sound and a disjoint pWF net $M$ is sub-sound and $p$ is a place in $N$ then $N\otimes_{p}M$ is sub-sound.\end{thm}
\begin{IEEEproof}
Assume that a tWF net $N$ is sub-sound and a pWF net $M$ is sub-sound. By definition of sub-soundness for tWF nets it follows that $\pc{N}$ is sub-sound. By Theorem~\ref{thm:place-subst-soundness-in-pWF} it follows that $\pc{N}\otimes_{p}M$ is sub-sound. By Lemma~\ref{lem:commut-pc-subst} it then holds that $\pc{N\otimes_{p}M}$ is sub-sound. Finally, by definition of sub-soundness for tWF nets, it follows that $N\otimes_{p}M$ is sub-sound.
\end{IEEEproof}

\begin{figure}
\centering{\includegraphics[width=0.3\textwidth]{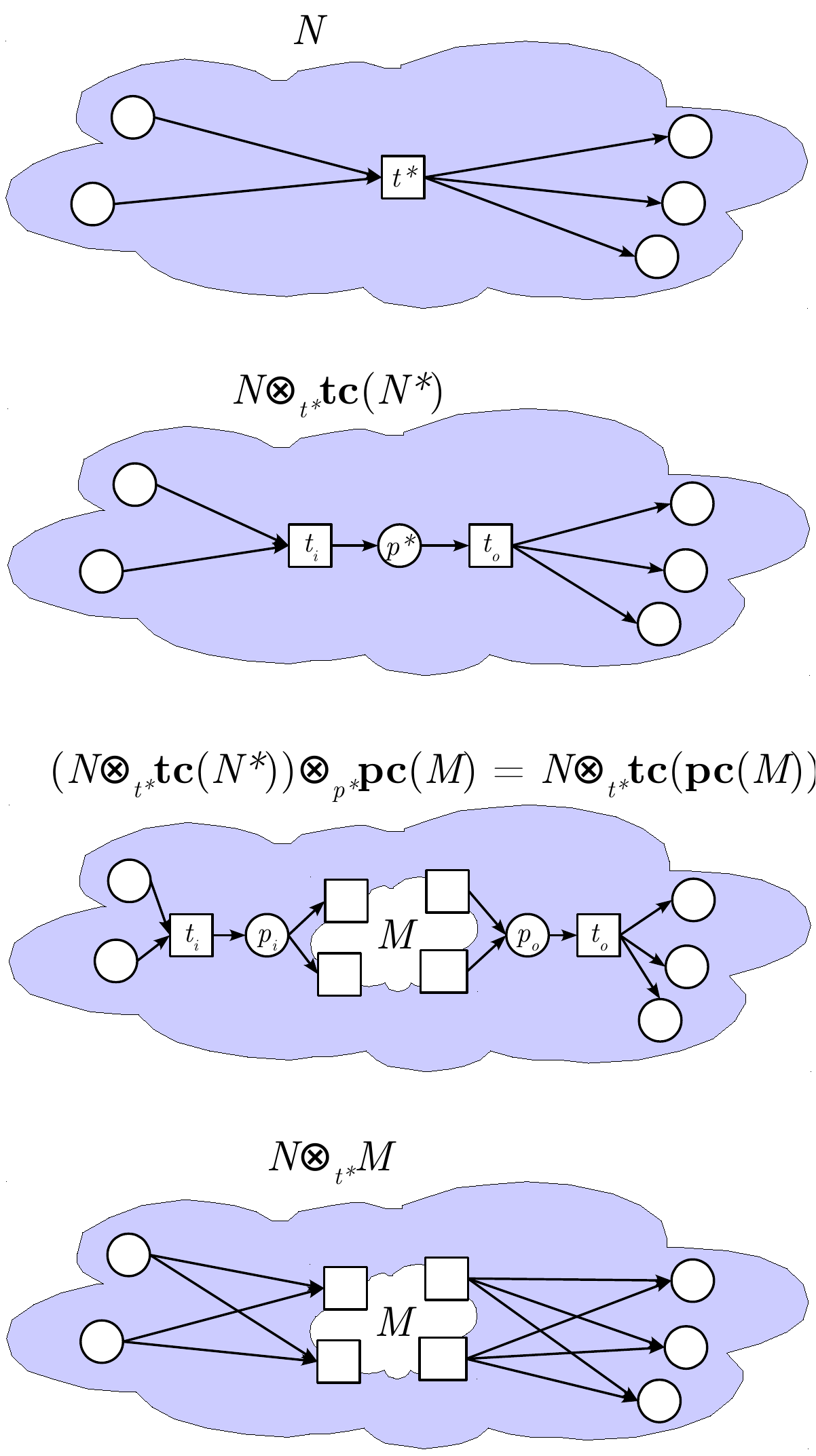}\caption{\label{fig:Trans-to-place-substitution5}Transforming transition substitution to place substitution}}
\end{figure}

We now proceed with showing that also transition substitution preserves sub-soundness. The proof strategy will be to show that this substitution is equivalent to a sequence of transformations with a place substitution as is illustrated in Figure~\ref{fig:Trans-to-place-substitution5}. The top net is the original net $N$ with transition $t^*$ that is to be replaced with net $M$, the result of which, i.e., $N\otimes_{t^*}M$, is shown in the bottom. The sequence of transformations with a place substitution is shown in between. In the second row we see $N\otimes_{t^*}\tc{N^{*}}$ where $N^*$ is a tWF consisting of transition completion of a single place $p^{*}$. As we show in Proposition~\ref{prop:subst-tpt-stays-subsound}, if $N$ is sub-sound, then $N\otimes_{t}^*\tc{N^{*}}$ also is sub-sound. Next, we see the result of substituting the place $p^{*}$ in $N\otimes_{t^*}\tc{N^{*}}$ with the pWF net $\pc{M}$. Finally, the input and output nodes introduced by the transition and place completions are removed, which also preserves sub-soundness as follows from Proposition~\ref{prop:trans-place-pair-removal} and Proposition~\ref{prop:place-trans-pair-removal}.

\begin{figure}
\centering{\includegraphics[width=0.375\textwidth]{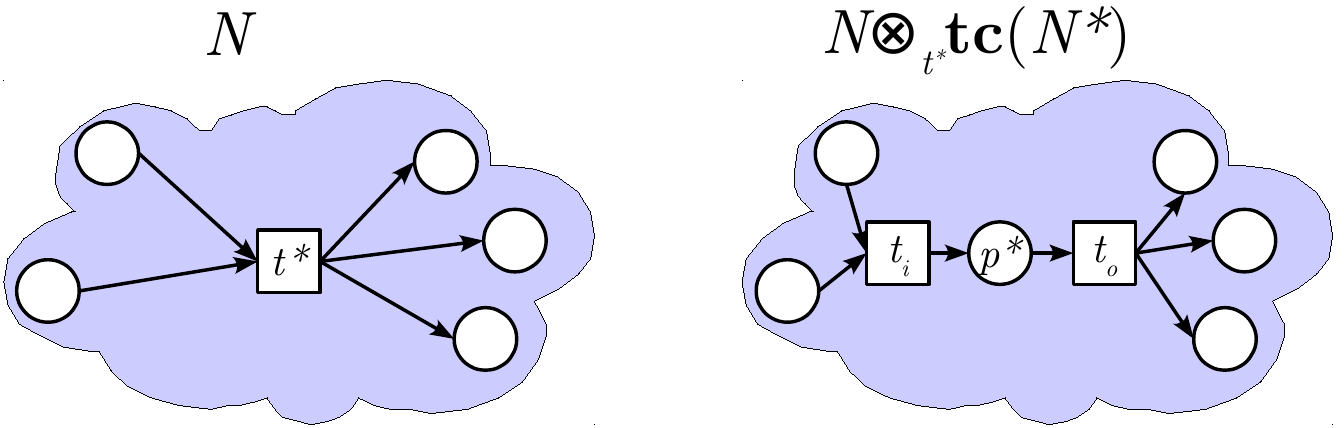}\caption{\label{fig:Trans-to-place-substitution4}Sequential transition substitution}}
\end{figure}

We begin with the lemma that shows that, see Figure~\ref{fig:Trans-to-place-substitution4}, if $N$ is sub-sound, then $N\otimes_{t}^*\tc{N^{*}}$ also is sub-sound.

\begin{prop}
\label{prop:subst-tpt-stays-subsound}If $N$ is a pWF net with a transition $t^{*}$ and $N^{*}$ a pWF net that consists of only a single place $p^{*}$, then $N\otimes_{t^{*}}\tc{N^{*}}$ is sub-sound if $N$ is sub-sound.
\end{prop}

\begin{IEEEproof}Let $M=N\otimes_{t^*}\tc{N^{*}}$. We define a relation $\sim\subseteq\mathbf{M}_{N}\times\mathbf{M}_{M}$ such that $m_{N}{\sim}m_{M}$ represents the fact that $m_{N}$ is the same as $m_{M}$ except that all (say $k$) tokens are removed from $p^{*}$ and $k$ tokens are added to each of $t_{o}\bullet_{M}$, or in other words, $t_{o}$ is fired $k$ times, where $t_{o}$ is the output transition added in $\tc{N^{*}}$. More formally: $m_{N}{\sim}m_{M}$ iff $m_{N}=m_{M}-[p^{*k}]+k.(t_{o}\bullet_{M})$ where $k=m_{M}(p^{*})$. 

It can then be shown that $\sim$ indeed defines a kind of bisimilarity, i.e., it holds that
\begin{description}
\item[(A*)] if $m_{N}\stackrel{\sigma}{\longrightarrow}_{N}m'_{N}$ and $m_{N}{\sim}m_{M}$, then $m_{M}\stackrel{*}{\longrightarrow}_{M}m'_{N}$, and
\item[(B*)] if $m \stackrel{\sigma}{\longrightarrow}_{M}m'_{M}$, then there is a marking $m_{N}$ such that $m\stackrel{*}{\longrightarrow}_{N}m_{N}$ and $m_{N}{\sim}m_{M}$.
\end{description}
Informally, this can be shown with induction on the length of $\sigma$. For the case of length 1 with transition $s$ we then distinguish for (A*) the cases where $s=t^*$ or not. Likewise for (B*) we distinguish the cases where $s=t_{i}$ or $s$ is a transition in $M$ not equal to $t_i$ nor $t_o$ (we will argue that with these assumptions $s \neq t_{o}$). We now proceed with showing that under the assumption of (A*) and (B*) we can indeed show that $M$ is sub-sound if $N$ is sub-sound.

\textbf{Proof that  $M$ is sub-sound if $N$ is sub-sound:} Note that, by construction of $M$, $N$ and $M$ have the same input set $I$ and output set $O$. Assume that N is sub-sound and that $k.I\stackrel{*}{\longrightarrow}_M(m_M+k'.O)$. By (B*) it follows that $k.I\stackrel{*}{\longrightarrow}_N m_N$ such that $m_N {\sim}(m_M + k'.O)$ that is $m_N = m_M + k'.O - [p^{*k''}] +k''.(t_o \bullet_M)$ where $k''=(m_{M}+k'.O)(p^{*})$. Since $p^{*} \notin O$ we can assume that $k''=m_{M}(p^{*})$ and that $m_N = (m_M - [p^{*k''}] +k''.(t_o \bullet_M)) + k'.O$, i.e., $k'.O \leq m_N$ and get $(m_N-k'.O) {\sim}m_M$. From the sub-soundness of $N$ it follows that $(m_N-k'.O)\stackrel{*}{\longrightarrow}_N (k-k').O$. Finally, by (A*) it follows that $m_M\stackrel{*}{\longrightarrow}_M (k-k').O$ which completes the proof of sub-soundness of M.

We will now formally show the missing (A*) and (B*). We start with the following facts:
\begin{description}

\item[(A)] If $m_{N}\stackrel{t}{\longrightarrow}_{N}m'_{N}$ and $m_{N}{\sim}m_{M}$, then $m_{M}\stackrel{*}{\longrightarrow}_{M}m'_{N}$. 

\item[(B)] If $m \stackrel{t}{\longrightarrow}_{M} m_{M}$, then there is an $m_{N}$ such that $m \stackrel{*}{\longrightarrow}_{N} m_{N}$ and $m_{N} {\sim}m_{M}$.

\end{description}

\textbf{Proof of claim (A):} If $m_{M}(p^{*})=k$, then we can fire $k$ times $t_{o}$ and so $m_{M}\stackrel{*}{\longrightarrow}_{M}m'_{M}=m_{M}-[p^{*k}]+k.(t_{o}\bullet_{M})$. Since $m_{N}{\sim}m_{M}$ we also have that $m_{N}=m_{M}-[p^{*k}]+k.(t_{o}\bullet_{M})$ and so $m'_{M}=m_{N}$, i.e., $m_{M}\stackrel{*}{\longrightarrow}_{M}m_{N}$. Either (i) $t\neq t^{*}$ or (ii) $t=t^{*}$. If (i), then by construction of $M$ we have $\bullet_N t = \bullet_M t$ and $t \bullet_N = t \bullet_M$ and so from $m_{N}\stackrel{t}{\longrightarrow}_{N}m'_{N}$ it follows $m_{N}\stackrel{t}{\longrightarrow}_{M}m'_{N}$. Thus we have shown that $m_{M}\stackrel{*}{\longrightarrow}_{M}m_{N}\stackrel{t}{\longrightarrow}_{M}m'_{N}$. If (ii), then by construction of $M$ we have $\bullet_N t = \bullet_M t_i$ and $t \bullet_N = t_o \bullet_M$ and so from $m_{N}\stackrel{t}{\longrightarrow}_{N}m'_{N}$ and the fact that $t_i \bullet_M = \bullet_M t_o$ it follows $m_{N}\stackrel{t_i, t_o}{\longrightarrow}_{M}m'_{N}$. Thus we have shown that $m_{M}\stackrel{*}{\longrightarrow}_{M}m_{N}\stackrel{t_i, t_o}{\longrightarrow}_{M}m'_{N}$.

\textbf{Proof of claim (B):} Because we assumed that $m$ is also a marking of N it holds that $m(p^*)= 0$, so either (i) $t\not\in\{t_{i},t_{o}\}$ or (ii) $t=t_{i}$. If (i), then by construction of $M$ we have $\bullet_N t = \bullet_M t$ and $t \bullet_N = t \bullet_M$ and so from $m \stackrel{t}{\longrightarrow}_{M} m_{M}$ it follows $m \stackrel{t}{\longrightarrow}_{N} m_{M}$. Of course $m_{M} {\sim}m_{M}$. If (ii), then by construction of $M$ we have $\bullet_N t^* = \bullet_M t_i$ and so from $m \stackrel{t_i}{\longrightarrow}_{M} m_{M}$ it follows $m \stackrel{t^*}{\longrightarrow}_{N} m_{N}$ for some $m_N$. We have $m_{N} = m - \bullet_N t^* + t^* \bullet_N = m - \bullet_M t_i + t_o \bullet_M$. On the other hand $m_M = m - \bullet_M t_i + t_i \bullet_M = m - \bullet_M t_i + p^*$. By combining these two we get $m_N = m_M - p^* + t_o \bullet_M$ and because $m(p^*)= 0$ we have $m_{M}(p^*)= 1$, so by definition $m_N {\sim}m_M$.

\textbf{Proof of claims (A*) and (B*):} The facts (A) and (B) can be generalized by induction on the length of $\sigma$ to show that (A*) if $m_{N}\stackrel{\sigma}{\longrightarrow}_{N}m'_{N}$ and $m_{N}{\sim}m_{M}$, then $m_{M}\stackrel{*}{\longrightarrow}_{M}m'_{N}$,
and (B*) if $m \stackrel{\sigma}{\longrightarrow}_{M} m_{M}$, then there is a marking $m_{N}$ such
that $m \stackrel{*}{\longrightarrow}_{N} m_{N}$ and $m_{N} {\sim}m_{M}$.
\end{IEEEproof}

We now proceed with propositions that show that the removal of $t_i$ and $p_i$  as well as $p_o$ and $t_o$ preserves sub-soundness. These results are similar to those of the abstraction rule of~\cite{Desel:2005:FCP:1205779}.

\begin{figure}
\begin{centering}
\includegraphics[width=0.375\textwidth]{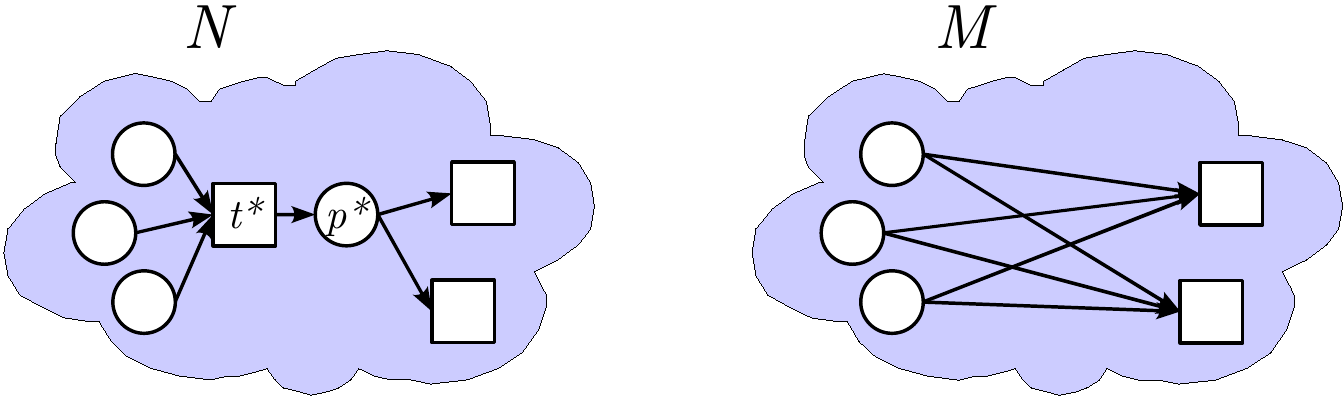}\caption{\label{fig:Transition-place-pair-removal}Transition-place pair removal}
\par\end{centering}
\end{figure}

\begin{prop}
\label{prop:trans-place-pair-removal}Let $N$ be a pWF net with transition $t^{*}$ and place $p^{*}$ such that $t^{*}\bullet_{N}= p^{*}$, $\bullet_{N}p^{*}=t^{*}$ and $p^{*}$ is not an input nor output place and there are no edges between $\bullet_{N}t^{*}$ and $p^{*}\bullet_{N}$. Furthermore, let $M$ be the pWF net that is obtained from $N$ if we remove $t^{*}$ and $p^{*}$ and add all the edges in $\bullet_{N}t^{*}\times p^{*}\bullet_{N}$ as illustrated in Figure~\ref{fig:Transition-place-pair-removal}. Then $M$ is sub-sound if $N$ is sub-sound.
\end{prop}

\begin{IEEEproof}
We define a similarity relation $\sim\subseteq\mathbf{M}_{N}\times\mathbf{M}_{M}$ such that $m_{N}{\sim}m_{M}$ represents the fact that $m_{M}$ is the same as $m_{N}$ except that all (say $k$) tokens are removed from $p^{*}$ and $k$ tokens are added to each of $\bullet t^{*}$, or in other words, $t^{*}$ is fired $k$ times in reverse. More formally: $m_{N}{\sim}m_{M}$ holds iff $m_{M}=m_{N}-[p^{*k}]+k.(\bullet_{N}t^{*})$ where $k=m_{N}(p^{*})$.

It can then be shown that $\sim$ defines a bisimilarity in the sense that:
\begin{description}
\item[ (D*) ] If $m_{N}\stackrel{\sigma}{\longrightarrow}_{N}m'_{N}$ and $m_{N}{\sim}m_{M}$, then there is a marking $m'_{M}$ such that $m_{M}\stackrel{*}{\longrightarrow}_{M}m'_{M}$ and $m'_{N}{\sim}m'_{M}$.
\item[(E*)] If $m_{M}\stackrel{\sigma}{\longrightarrow}_{M}m'_{M}$ and $m_{N}{\sim}m_{M},$ then there is a marking $m'_{N}$ such that $m_{N}\stackrel{*}{\longrightarrow}_{N}m'_{N}$ and $m'_{N}{\sim}m'_{M}$.
\item[(F*)] If $m \stackrel{*}{\longrightarrow}_{M} m_M$, $m \stackrel{\sigma}{\longrightarrow}_{N} m_N$, $m_N {\sim}m_M$ and $m_N(p^*) > 0$ then from $\sigma$ we can construct $\sigma'$ by removing the last $m_N(p^*)$ occurrences of $t^*$ and get $m \stackrel{\sigma'}{\longrightarrow}_{N} m'_N$, $m'_N {\sim}m_M$ and $m'_N(p^*) = 0$.
\end{description}
Informally this can be shown with induction on the length of $\sigma$. In the case of a single transition $t$ in $\sigma$ we distinguish for (D*) the cases where $t=t^{*}$ and if not then $p^{*}\in\bullet_{N}t$ or not. Likewise, for (E*) we distinguish the cases where $p^{*}\in\bullet_{N}t$ or not. Finally, for (F*) we observe that all $m_N(p^*)$ tokens in $p^*$ had to placed there by $t^*$ during $\sigma$ and that the last of those tokens is not needed by the following transitions of $\sigma$.

Now, using (D*), (E*) and (F*), we show that $M$ is sub-sound if $N$ is sub-sound. Note that, by construction, $N$ and $M$ have the same input set $I$ and output set $O$, and that $O{\sim}O$.

\textbf{Proof that $M$ is sub-sound if $N$ is sub-sound:} Assume that $k.I\stackrel{*}{\longrightarrow}_{M}(m_{M}+k'.O)$. By (E*) it follows that $k.I\stackrel{*}{\longrightarrow}_{N}m_{N}$ such that $m_{N} {\sim}(m_{M}+k'.O)$. By (F*) $k.I\stackrel{*}{\longrightarrow}_{N}m'_{N}$ where $m'(p^*)=0$ and $m'_{N} {\sim}(m_{M}+k'.O)$. By definition of $\sim$ the last two give $m'_{N} = (m_{M}+k'.O)$, i.e., $k.I\stackrel{*}{\longrightarrow}_{N} (m_{M}+k'.O)$. Now by sub-soundness of $N$ it follows that $m_{M} \stackrel{*}{\longrightarrow}_{N} (k-k').O)$. By (D*) $m_{M} \stackrel{*}{\longrightarrow}_{M} m'_M$ and $(k-k').O {\sim}m'_M$. Since $(k-k').O(p^*) = 0$ we get $m'_M=(k-k').O$.

We now will formally show the missing (D*), (E*) and (F*). We start with the following auxiliary claims: 
\begin{description}

\item[(A)] If $m_{N}{\sim}m_{M}$ and $t$ such that $t\neq t^{*}$ and $\bullet_{N}t\leq m_{N}$ then $\bullet_{M}t\leq m_{M}$. 

\item[(B)] If $\bullet_{M}t \leq m$ then there is an $m_N {\sim}m$ such that $m \stackrel{*}{\longrightarrow}_{N} m_{N}$, $\bullet_{N}t \leq m_{N}$. 

\item[(C)]  If $m_{N}{\sim}m_{M}$ and $m_{N}\stackrel{t}{\longrightarrow}_{N} m'_{N}$ and $m_M \stackrel{t}{\longrightarrow}_M m_{M}'$ then $m'_{N}{\sim}m'_{M}$. 

\end{description}

\textbf{Proof of claim (A):} Assume that $m_{N}{\sim}m_{M}$, $t\neq t^{*}$ and $\bullet_{N}t\leq m_{N}$. From $m_{N}{}{\sim}m_{M}$ it follows that $m_{M}=m_{N}-[p^{*k}]+k.(\bullet_{N}t^{*})$ where $k=m_{N}(p^{*})$. Consider the case where $p^{*}\in\bullet_{N}t$. Then $\bullet_{M}t=\bullet_{N}t-[p^{*}]+\bullet_{N}t^{*}\leq m_{N}-[p^{*k}]+k.\bullet_{N}t^{*}=m_{M}$ where the first equality follows from the definition of $M$ and the inequality from the observation that in this case $k \geq 1$. Consider the other case where $p^{*}\notin \bullet_{N}t$. Here from $\bullet_N t \leq m_N$ it follows that $\bullet_{N}t\leq m_{N}-[p^{*k}]+k.\bullet_{N}t^{*}$ and we get $\bullet_{M}t=\bullet_{N}t\leq m_{N}-[p^{*k}]+k.\bullet_{N}t^{*}=m_{M}$. 

\textbf{Proof of claim (B):} Assume that $\bullet_{M}t \leq m$. Consider the case where $p^{*}\notin\bullet_{N}t$. Then $\bullet_{N}t=\bullet_{M}t\leq m$ and so we can take $m_{N}=m$. Consider the other case where $p^{*}\in\bullet_{N}t$. By the construction $\bullet_{N}t^{*} \leq \bullet_{M}t \leq m$, i.e., $t$ has to be enabled in $N$. Let $m_N$ be a marking such that $m \stackrel{t^*}{\longrightarrow_{N}} m_{N}$ that is $m_N = m - \bullet_{N}t^{*} + p^*$. Since $\bullet_{N}t = \bullet_{M}t - \bullet_{N}t^{*} + p^*$  this implies that $\bullet_{N}t \leq m_N$.

\textbf{Proof of claim (C):} Assume that $m_{N} {\sim}m_{M}$ and $m_{N}\stackrel{t}{\longrightarrow}_{N} m'_{N}$ and $m_{M}\stackrel{t}{\longrightarrow}_{M} m'_{M}$. Because $m_{N}{\sim}m_{M}$, $m_M = m_{N}-[p^{*k}]+k.(\bullet_{N}t^{*})$ where $k=m_{N}(p^{*})$. Because $m_{N}\stackrel{t}{\longrightarrow}_{N} m'_{N}$, $m'_N = m_N - \bullet_N t + t\bullet_{N}$. Because $m_{M}\stackrel{t}{\longrightarrow}_{M} m'_{M}$, $m'_M = m_M - \bullet_M t + t\bullet_M$. By construction and because $t \neq t^{*}$, $t\bullet_{M}=t\bullet_{N}$. Now either (i) $p^{*}\in\bullet_{N}t$ or (ii) $p^{*} \notin \bullet_{N}t$. If (i) then $k \geq 1$ and $\bullet_{M}t=\bullet_{N}t-[p^{*}]+\bullet_{N}t^{*}$. It follows that $m'_M = (m_N-[p^{*k}]+k.(\bullet_{N}t^{*}) - (\bullet_{N}t-[p^{*}]+\bullet_{N}t^{*}) + t\bullet_{N}) = m_N -[p^{*(k-1)}] + (k-1).(\bullet_{N}t^{*} - \bullet_{N}t + t\bullet_{N} = m_N - \bullet_{N}t + t\bullet_{N} - [p^{*(k-1)}] + (k-1).(\bullet_{N}t^{*} = m'_N - [p^{*(k-1)}] + (k-1).(\bullet_{N}t^{*}$ with $m'_N(p^*)=k-1$ since $p^* \in \bullet_N t$ and $p^* \notin t\bullet_N$. Thus $m'_N {\sim}m'_M$. Consider the other case (ii) where $p^{*} \notin \bullet_{N}t$. Then $\bullet_M t = \bullet_N t$ and therefore $m'_{M}=(m_N-[p^{*k}]+k.(\bullet_{N}t^{*})-\bullet_{N}t+t\bullet_{N} = m_N-\bullet_{N}t+t\bullet_{N}-[p^{*k}]+k.(\bullet_{N}t^{*}) = m'_N - [p^{*k}] + k.(\bullet_{N}t^{*})$ with $m'_{N}(p^{*})=k$, since $p^{*} \notin \bullet_{N}t$ and $p^{*} \notin t\bullet_{N}$. Thus $m'_{N}{\sim}m'_{M}$. 

We then show the claims that concern the cases of (D*), (E*) and (F*) where $\sigma$ is of length 1:
\begin{description}

\item[(D)] If $m_{N}\stackrel{t}{\longrightarrow}_{N}m'_{N}$ and $m_{N}{\sim}m_{M}$ then there is a marking $m'_{M}$ such that $m_{M}\stackrel{*}{\longrightarrow}_{M}m'_{M}$ and $m'_{N}{\sim}m'_{M}$.

\item[(E)] If $m \stackrel{t}{\longrightarrow}_{M} m_{M}$ then there is a marking $m_{N}$ such that $m \stackrel{*}{\longrightarrow}_{N} m_{N}$ and $m_{N}{\sim}m_{M}$. 

\item[(F)] If $m \stackrel{*}{\longrightarrow}_{M} m_M$, $m \stackrel{\sigma}{\longrightarrow}_{N} m_N$, $m_N {\sim}m_M$ and $m_N(p^*) > 0$ then from $\sigma$ we can construct $\sigma'$ by removing the last occurrence of $t^*$ and get $m \stackrel{\sigma'}{\longrightarrow}_{N} m'_N$, $m'_N {\sim}m_M$ and $m'_N(p^*) = m_N(p^*)-1$.

\end{description}

\textbf{Proof of claim (D):} Assume that $m_{N}\stackrel{t}{\longrightarrow}_{N}m'_{N}$ and $m_{N}{\sim}m_{M}$. Now either (i) $t=t^{*}$ or (ii) $t\neq t^{*}$. If (i) then $m'_{N}{\sim}m_{M}$ and so we can take $m'_{M}=m_{M}$. Consider the case (ii) where $t\neq t^{*}$. By (A) it then holds that $t$ is enabled in $m_{M}$ for $M$, and so $m_{M}\stackrel{t}{\longrightarrow}_{M}m'_{M}$ for some $m'_{M}$. By (C) it then follows that $m'_{N}{\sim}m'_{M}$.

\textbf{Proof of claim (E):} Assume that $m \stackrel{t}{\longrightarrow}_{M} m_{M}$. By (B) there is an $m_{N}$ such that $m \stackrel{*}{\longrightarrow}_{N} m_{N}$, $\bullet_{N}t \leq m_{N}$ and $m_{N} {\sim}m$. Since $\bullet_{N}t \leq m_{N}$ it holds that $m_{N} \stackrel{t}{\longrightarrow}_{N} m'_{N}$ for some $m'_{N}$. By (C) it then follows that $m'_{N} {\sim}m$.

\textbf{Proof of claim (F):} Assume that $m \stackrel{*}{\longrightarrow}_{M} m_M$, $m \stackrel{\sigma}{\longrightarrow}_{N} m_N$, $m_N {\sim}m_M$ and $m_N(p^*) > 0$. Since $m$ is a marking of both $N$ and $M$, it does not place any tokens in $p^*$ which is not present in $M$. So all $m_N(p^*)$ tokens in $p^*$ had to be placed there during $\sigma$ by firing $t^*$, which is the only transition that can do that, and the token placed there as last is not needed by the following transitions of $\sigma$. This is due to the fact that we do not distinguish individual tokens of a place and without the loss of generality we can assume that places act as FIFO queues for tokens. Thus a valid firing sequence $\sigma'$  can be constructed from $\sigma$ by removing the last occurrence of $t^*$. Let $m \stackrel{\sigma'}{\longrightarrow}_{N} m'_N$. By the definition of $\sigma'$ it holds that $m'_N = m_N - p^* + \bullet_N t^*$. It follows that $m'_N {\sim}m_M$ and $m'_N(p^*) = m_N(p^*)-1$.

Finally, we now turn to the proofs of (D*), (E*) and (F*):

\textbf{Proofs of claims (D*), (E*) and (F*):} With induction on the length of $\sigma$ it follows form (D) that (D*) if $m_{N}\stackrel{\sigma}{\longrightarrow}_{N}m'_{N}$ and $m_{N}{\sim}m_{M}$ then there is a marking $m'_{M}$ such that $m_{M}\stackrel{*}{\longrightarrow}_{M}m'_{M}$ and $m'_{N}{\sim}m'_{M}$. Likewise it follows from (E) that (E*) if $m_{M}\stackrel{\sigma}{\longrightarrow}_{M}m'_{M}$ and $m_{N}{\sim}m_{M}$ then there is a marking $m'_{N}$ such that $m_{N}\stackrel{*}{\longrightarrow}_{N}m'_{N}$ and $m'_{N}{\sim}m'_{M}$. Finally it follows from (F) that F*) if $m \stackrel{*}{\longrightarrow}_{M} m_M$, $m \stackrel{\sigma}{\longrightarrow}_{N} m_N$, $m_N {\sim}m_M$ and $m_N(p^*) > 0$ then from $\sigma$ we can construct $\sigma'$ by removing the last $m_N(p^*)$ occurrences of $t^*$ and get $m \stackrel{\sigma'}{\longrightarrow}_{N} m'_N$, $m'_N {\sim}m_M$ and $m'_N(p^*) = 0$.
\end{IEEEproof}

\begin{figure}
\centering{}\includegraphics[width=0.375\textwidth]{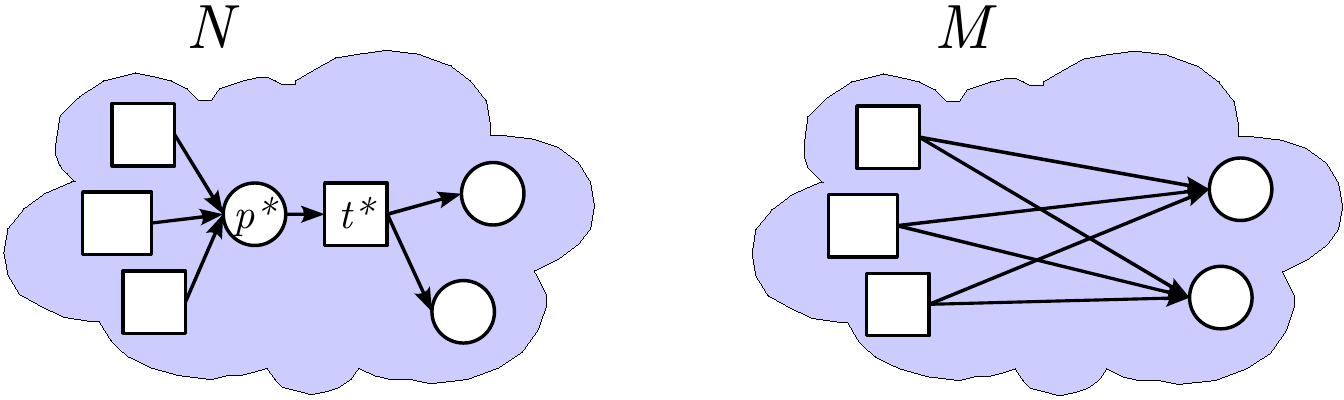}\caption{\label{fig:Place-transition-pair-removal}Place-transition pair removal}
\end{figure}

\begin{prop}
\label{prop:place-trans-pair-removal}Let $N$ be a pWF net with place $p^{*}$ and transition $t^{*}$ such that $p^{*}\bullet_{N}=t^{*}$, $\bullet_{N}t^{*}=p^{*}$ and $p^{*}$ is not an input nor output place and there are no edges between $\bullet_{N}p^{*}$ and $t^{*}\bullet_{N}$. Furthermore, let $M$ be the pWF net that is obtained from $N$ if we remove $p^{*}$ and $t^{*}$ and add all the edges in $\bullet_{N}p^{*}\times t^{*}\bullet_{N}$ as illustrated in Figure~\ref{fig:Place-transition-pair-removal}. Then $M$ is sub-sound if $N$ is sub-sound.
\end{prop}
\begin{IEEEproof}
The proof proceeds analogously to that of the preceding Proposition~\ref{prop:trans-place-pair-removal} with the relation $\sim\subseteq\mathbf{M}_{N}\times\mathbf{M}_{M}$ redefined such that $m{\sim}m'$ iff $m'=m-[p^{*k}]+k.(t^{*}\bullet_{N})$ where $k=m(p^{*})$.
\end{IEEEproof}

We are now ready to prove that sub-soundness is preserved by transition substitution.

\begin{thm}
\label{thm:trans-subst-soundness-in-pWF}If a pWF net $N$ is sub-sound and a disjoint tWF net $M$ is sub-sound and $t^*$ is a transition in $N$, then $N\otimes_{t^*}M$ is sub-sound.\end{thm}

\begin{IEEEproof}Let $N=(P,T,F,I,O)$ be a sub-sound pWF net containing a transition $t$, and $M=(P',T',F',I',O')$ a sub-sound tWF net. Furthermore, let $N^{*}$ be a pWF net consisting of a single new place $p^* \notin P \cup P'$. We will construct $N\otimes_{t^*}M$ by a sequence of transformations and substitutions where the sub-soundness of the result of each step will follow from the sub-soundness of the nets used as components.

Consider the sequence of transformation in Figure~\ref{fig:Trans-to-place-substitution5}. In the top we start with $N$ which by assumption is sub-sound. We first substitute $t^*$ with $\tc{N^*}$ and get $N\otimes_{t^*}\tc{N^{*}}$ which by Proposition~\ref{prop:subst-tpt-stays-subsound} is sub-sound if $N$ is sub-sound. Then we substitute $p^*$ with $\pc{M}$ and get  $(N\otimes_{t^*}\tc{N^{*}})\otimes_{p^{*}}\pc{M}$. Here the sub-soundness follows from Theorem~\ref{thm:place-subst-soundness-in-pWF} and the fact that a place completion a sub-sound tWF net is sub-sound by definition. Finally we remove nodes $t_i$ and $p_i$ as well as $p_o$ and $t_o$ by applying Propositions~\ref{prop:trans-place-pair-removal} and~\ref{prop:place-trans-pair-removal} respectively. This concludes the proof that the resulting net $N\otimes_{t^*}M$ is sub-sound.
\end{IEEEproof}

\begin{thm}
\label{thm:trans-subst-soundness-in-tWF}If a tWF net $N$ is sub-sound
and a disjoint tWF net $M$ is sub-sound and $t$ is a transition
in $N$ then $N\otimes_{t}M$ is sub-sound.\end{thm}
\begin{IEEEproof}
Assume that $N$ is sub-sound tWF net with a transition $t$ and $M$ a sub-sound tWF net. By Theorem~\ref{thm:trans-subst-soundness-in-pWF} it follows that $\pc{N}\otimes_{t}M$ is sub-sound. Since by Lemma~\ref{lem:commut-pc-subst} it holds that $\pc{N\otimes_{t}M}=\pc{N}\otimes_{t}M$, it follows that $\pc{N\otimes_{t}M}$ is sub-sound. By definition of sub-soundness of tWF nets it then holds that $N\otimes_{t}M$ is sub-sound.
\end{IEEEproof}
\begin{cor}
\label{cor:subst-subsound}If $N$ and $M$ are disjoint sub-sound
WF nets and $n$ is a node in $N$ then $N\otimes{}_{n}M$ (if defined) is a sub-sound WF net.\end{cor}
\begin{IEEEproof}
This follows from the fact that Theorem~\ref{thm:place-subst-soundness-in-pWF}, Theorem~\ref{thm:place-subst-soundness-in-tWF}, Theorem~\ref{thm:trans-subst-soundness-in-pWF} and Theorem~\ref{thm:trans-subst-soundness-in-tWF} cover all possible combinations of $N$ and $M$ being pWF nets or tWF nets.
\end{IEEEproof}

\section{Sub-soundness of AND-OR nets}\label{sec:subsoundness_AND-OR}
In this section we show that all AND-OR nets are sub-sound. First we show that the AND nets and OR nets from which AND-OR nets are generated are.

\begin{thm}
\label{thm:one-one-pOR-subsound}Every one-input one-output pOR net
is sub-sound.\end{thm}
\begin{IEEEproof}
Since in OR nets transitions cannot have multiple input/output places it can be shown by induction on the length of $\sigma$ that (A) if $|m|=k$ and $m\stackrel{\sigma}{\longrightarrow}m'$ then $|m'|=k$. Let $I_{N}=\{p_{i}\}$ and $O_{N}=\{p_{o}\}$. For each place $p$ in a pOR net $N$ it holds that $[p_{i}]\stackrel{*}{\longrightarrow}[p]$ and $[p]\stackrel{*}{\longrightarrow}[p_{o}]$ since there must be paths from $p_{i}$ to $p$ and from $p$ to $p_{o}$ and each transition in those paths has one input edge and one output edge. Thus, it also follows that (B) if $|m|=k$, then $k.[p_{i}]\stackrel{*}{\longrightarrow}m$ and $m\stackrel{*}{\longrightarrow}k.[p_{o}]$.

We now show the sub-soundness requirement. Assume that $k.I_{N}\stackrel{*}{\longrightarrow}(m+k'.O_{N})$. Since $|k.I_{N}|=|k.[p_{i}]|=k.|[p_{i}]|=k$ it follows by (A) that $|m+k'.O_{N}|=k$. Since $|m+k'.O_{N}|=|m|+|k'.O_{N}|$ and $|k.O_{N}|=|k.[p_{o}]|=k * |[p_{o}]|=k$ it follows that $|m|=k-k'$. By (B) it then follows that $m\stackrel{*}{\longrightarrow}(k-k').[p_{o}]=(k-k').O_{N}$.\end{IEEEproof}

\begin{thm}
\label{thm:tOR-subsound}Every tOR net is sub-sound.\end{thm}
\begin{IEEEproof}
Consider a tOR net $N$. By the definition of {*}-soundness of tWF nets it holds that $N$ is {*}-sound if $\pc{N}$ is {*}-sound. Observe that $\pc{N}$ is an one-input one-output pAND net, because $N$ by definition it does not have and incoming edges of the input places nor outgoing edges of the output places. By Theorem~\ref{thm:one-one-pOR-subsound} it holds that $\pc{N}$ is sub-sound and therefore {*}-sound. By Lemma~\ref{lem:tWF-subsound-is-starsound} it follows that $N$ is sub-sound.\end{IEEEproof}

\begin{thm}
\label{thm:pAND-subsound}Every pAND net is sub-sound.\end{thm}
\begin{IEEEproof}
Consider a pAND net $N$. Thanks to the limit on the number of incoming edges of the input places and outgoing edges of the output places in the definition of AND net $\tc{N}$ is a one-input one-output tAND net. Also $\tc{N}$ does not have incoming edges of the input transition nor outgoing edges of the output transition. By Theorem~17 in~\cite{DBLP:conf/apn/HeeSV03} we get that $\pc{\tc{N}}$ is {*}-sound which by definition gives proves that $\tc{N}$ is {*}-sound. By Theorem~\ref{thm:tc-soundness} it follows that $N$ is sub-sound and therefore {*}-sound. Therefore by Lemma~\ref{lem:sub-is-star-if-no-out} it follows that $N$ is sub-sound.\end{IEEEproof}

\begin{thm}
\label{thm:one-one-tAND-subsound}Every one-input one-output tAND net is sub-sound.\end{thm}
\begin{IEEEproof}
Consider a one-input, one-output tAND net $N$. It's input transition cannot have incoming edges nor its output transition cannot have outgoing edges, since those would have to introduce cycles. By Theorem~\ref{thm:pAND-subsound} it follows that $pc(N)$ is sub-sound, so also {*}-sound, and thus $N$ in {*}-sound. Therefore by Lemma~\ref{lem:tWF-subsound-is-starsound} it follows that $N$ is sub-sound.\end{IEEEproof}

\begin{cor}
All AND-OR nets are sub-sound.\end{cor}
\begin{IEEEproof}
By Theorem~\ref{thm:one-one-pOR-subsound}, Theorem~\ref{thm:tOR-subsound},
Theorem~\ref{thm:one-one-tAND-subsound} and Theorem~\ref{thm:pAND-subsound}
the initial nets are all sub-sound, and by Corollary~\ref{cor:subst-subsound}
substitution preserves sub-soundness.
\end{IEEEproof}

\section{Future Research}\label{sec:future_research}

The class of AND-OR nets can be researched further in several ways.
One direction could be to attempt to characterize the class in terms
of syntactic and semantic properties. As was shown all the nets in
it are sound, even sub-sound, and it is also not hard to see that
they are all free-choice nets, but it certainly not true that the
class contains all sub-sound free-choice nets as is show in Theorem~\ref{thm:Not-all-free-choice}.
So it remains open which semantic property characterizes the AND-OR
nets.

\begin{center}\begin{figure}
\centering{}\includegraphics[width=0.4\textwidth]{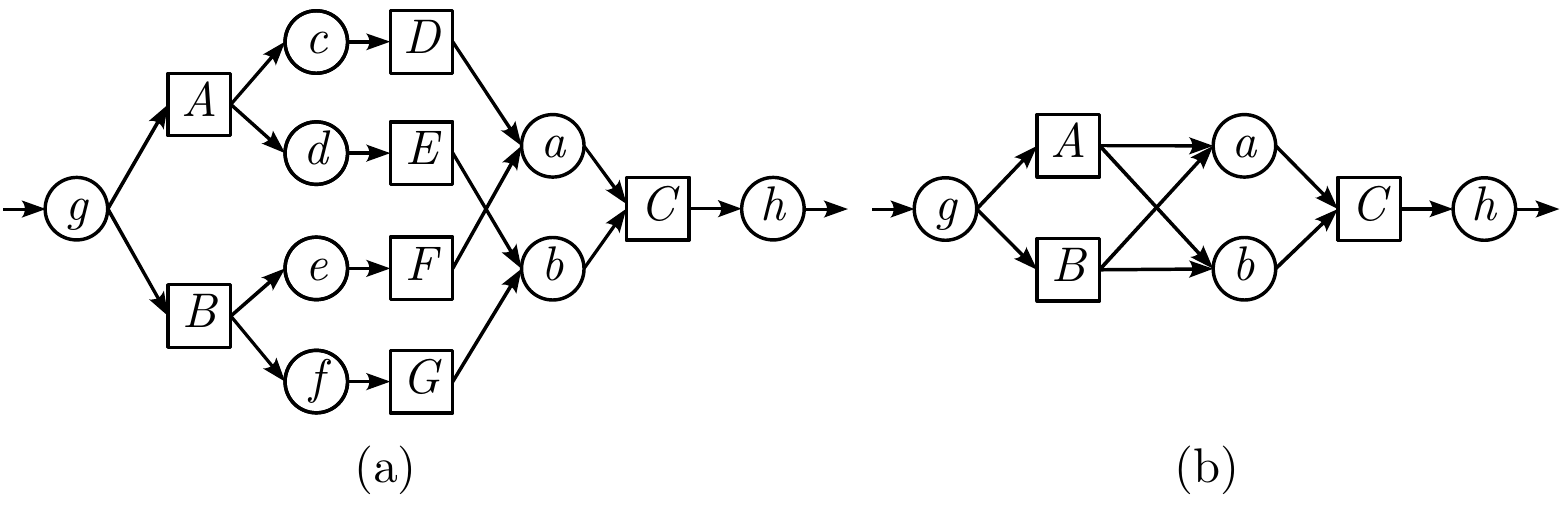}\caption{\label{fig:Counterexample-AND-OR-incomplete}Counterexample for the
completeness of AND-OR nets}
\end{figure}\end{center}

\begin{thm}
\label{thm:Not-all-free-choice}Not all free-choice sub-sound workflow
nets are AND-OR nets.\end{thm}
\begin{IEEEproof}
The counterexample is given in Figure~\ref{fig:Counterexample-AND-OR-incomplete}
(a) (taken from \cite{DBLP:conf/apn/HeeSV03}).
\end{IEEEproof}

Another potential research direction is the extension of the class
by introducing new forms of substitution that still can be considered
hierarchical. For example, it might be allowed that not only substitute
nodes but also edges: an edge from a place to a transition could be
replaced with a workflow net starting with a single place and ending
with a single transition. In general such substitutions do not preserve
sub-soundness, but they can be syntactically restricted such that
they do. To illustrate, such substitutions could be used to generate
Figure~\ref{fig:Counterexample-AND-OR-incomplete} (a) from the AND-OR
net in Figure~\ref{fig:Counterexample-AND-OR-incomplete} (b) by
substituting the edges $(A,a)$, $(A,b)$, $(B,a)$ and $(B,b)$.

Yet another possible generalization can be achieved by weakening the
requirement that a substitution links all the input and output nodes
in the same way. For example, it could be allowed that a transition
is replaced with a tAND net with a single input transition and several
output transitions such that (1) each output transition in the tAND
net is linked to at least one place in the postset of the replaced
transition and (2) each place in the postset of the replaced transition
is linked with exactly one output transition in the tAND net. Also
this would allow us to generate Figure~\ref{fig:Counterexample-AND-OR-incomplete}
(a) from the AND-OR net in Figure~\ref{fig:Counterexample-AND-OR-incomplete}
(b) by substituting the transitions $A$ and $B$.

\section{Conclusions}

We have investigated an approach for generating sound workflow nets in a structured way. This approach is based on the notion of a substitution of one node by a workflow net with input and output nodes being of the same type as the substituted node. The substituted nets can have multiple inputs and outputs, which is an extension of the previously considered substitutions and allows to generate a larger class of nets. We have identified a specific notion of soundness that is preserved by such substitutions and which allows to show that the generated nets are indeed sound.


\bibliographystyle{plain}
\bibliography{refinement}

\end{document}